\documentclass[a4paper]{article}

\usepackage{PRIMEarxiv}
\usepackage[utf8]{inputenc}
\usepackage[T1]{fontenc}
\usepackage{url}            
\usepackage{booktabs}       
\usepackage{amsfonts}       
\usepackage{nicefrac}       
\usepackage{microtype}      
\usepackage{lipsum}
\usepackage{fancyhdr}       
\usepackage{graphicx}       
\usepackage{multirow}
\usepackage{subfigure}
\usepackage{stmaryrd}
\usepackage[version=4]{mhchem}
\usepackage[export]{adjustbox}
\usepackage{bbold}
\usepackage{eurosym}

\usepackage[numbers]{natbib}  
\usepackage{hyperref}
\hypersetup{
	colorlinks=true,       
	linkcolor=blue,        
	citecolor=blue,        
	filecolor=blue,        
	urlcolor=blue          
}

\usepackage{doi}
\renewcommand{\doi}[1]{\href{https://doi.org/#1}{doi:#1}}

\title{A Robust Sustainability Assessment Methodology for Aircraft Parts: Application to a Fuselage Panel}

\author{
	Aikaterini A. Anagnostopoulou \\
	Laboratory of Technology \& Strength of Materials \\
	Department of Mechanical Engineering \& Aeronautics \\
	University of Patras, 26500 Patras, Greece \\
	\texttt{aikat.anagnostopoulou@upnet.gr} \\
	\And
	Dimitris G. Sotiropoulos \\
	Department of Electrical and Computer Engineering \\
	University of Peloponnese, 26334 Patras, Greece \\
	\texttt{dg.sotiropoulos@uop.gr} \\
	\And
	Konstantinos I. Tserpes \\
	Laboratory of Technology \& Strength of Materials \\
	Department of Mechanical Engineering \& Aeronautics \\
	University of Patras, 26500 Patras, Greece \\
	\texttt{kitserpes@upatras.gr}
}

\date{}

\begin{document}
	\maketitle
	
	\begin{abstract}
		The paper presents a cradle-to-gate sustainability assessment methodology specifically designed to evaluate aircraft components in a robust and systematic manner. This methodology integrates multi-criteria decision-making (MCDM) analysis across ten criteria, categorized under environmental impact, cost, and performance. Environmental impact is analyzed through life cycle assessment and cost through life cycle costing, with both analyses facilitated by SimaPro software. Performance is measured in terms of component mass and specific stiffness. The robustness of this methodology is tested through various MCDM techniques, normalization approaches, and objective weighting methods. To demonstrate the methodology, the paper assesses the sustainability of a fuselage panel, comparing nine variants that differ in materials, joining techniques, and part thicknesses. All approaches consistently identify thermoplastic CFRP panels as the most sustainable option, with the geometric mean aggregation of weights providing balanced criteria consideration across environmental, cost, and performance aspects. The adaptability of this proposed methodology is illustrated, showing its applicability to any aircraft component with the requisite data. This structured approach offers critical insights to support sustainable decision-making in aircraft component design and procurement.
	\end{abstract}
	

	\newpage
	
	\tableofcontents
	
	\newpage

\section{Introduction}
Greenhouse gas emissions from aviation have risen significantly over the past three decades. Although aviation, along with shipping, accounts for about 4\% of the European Union's total emissions, it remains the fastest-growing source of emissions. The primary driver behind this increase is the growth in air traffic. Efforts to reduce emissions in the aviation sector have only recently gained traction on a global scale. The EU aims to reduce emissions by 55\% by 2030 and achieve net-zero emissions by 2050. To meet these targets, the EU is currently implementing several strategies, including the inclusion of aviation in the emissions trading scheme, a revision of this scheme for aviation, and proposals for more sustainable fuels and aircraft technologies~\cite{Vogiantzi2023}. In the field of aviation, relevant research is coordinated by the Clean Aviation Joint Undertaking, established with the primary mission of developing disruptive aircraft technologies to support the European Green Deal and achieve climate neutrality by 2050. One of the main technologies emphasized by Clean Aviation is ``Green manufacturing and assembly, end-to-end and eco-design.'' According to the Strategic Research and Innovation Agenda of Clean Aviation, ``A key enabler of green manufacturing and eco-design is sustainability assessment.''

In contrast to other engineering sectors, such as automotive~\cite{Drohomeretski2015}, where sustainability practices are already integrated into structural design, relatively few studies have focused on sustainability assessments in aviation. Specifically, a search in Scopus for ``Sustainability'' AND ``Aircrafts'' reveals only 21 relevant journal articles. For instance, \cite{Pinheiro2020} outlines an approach for modeling future aircraft technologies and provides an overview of existing methods and challenges in assessing sustainability for current technologies. Reference~\cite{Swastanto2024} explores the integration of sustainability practices within aircraft maintenance, repair, and overhaul (MRO) companies. Meanwhile, \cite{Filippatos2024} proposes a sustainability-based design process that evaluates sustainability using an index encompassing technology, environmental, economic, and circular economy aspects, first introduced in~\cite{Katsiropoulos2020}.

Research has extensively examined how propulsion systems and fuel types impact aircraft sustainability. In~\cite{Balli2023,Akdeniz2023}, turbojet engines fueled by kerosene and bio-based alternatives underwent comprehensive energy, exergy, thermo-ecological, environmental, enviroeconomic, and sustainability assessments. Similarly,~\cite{Akdeniz2022} investigates the potential benefits of hydrogen fuel in enhancing sustainability and environmental performance in a medium-scale turboprop engine. Sustainability and economic efficiencies of electric vehicles and aircraft are compared in~\cite{Markowska2023}. In~\cite{Barbosa2023}, a hybrid simulation approach assesses the sustainability of make-to-order supply chains. Further research, such as that by Karpuk et al.~\cite{Karpuk2022}, illustrates the effect of new airframe and propulsion technologies on the sustainability of future medium-range jets. In~\cite{Aygun2021}, a thermodynamic, environmental, and sustainability analysis is conducted on the PW4000 engine across eight long-haul flight phases.

Additionally,~\cite{Fera2020} proposes an aircraft design process that incorporates sustainability throughout its service life, introducing a green index that combines maintenance costs with an environmental parameter. Reference~\cite{Cardeal2020} presents a model to support the design and evaluation of sustainable business models, while~\cite{Lin2019} examines 3D printing's role in enhancing sustainability within an MRO setting. In~\cite{Baklacioglu2018}, an exergetic metaheuristic design for a business jet aircraft predicts an exergetic sustainability index and an environmental effect factor using artificial neural networks for various flight phases. A sustainability index for an aircraft manufacturing company using a Fuzzy Best-Worst decision-making approach is introduced in~\cite{Raj2018}. Exergy modeling to evaluate the sustainability level of high by-pass turbofan engines in commercial aircraft is presented in~\cite{Balli2017}.

Innovative concepts and methodologies continue to emerge. In~\cite{Somerville2016}, box wing aircraft technology is introduced alongside an overview of ongoing research efforts, while~\cite{Sabaghi2015} describes an efficient fuzzy method for assessing sustainability in dismantling strategies, considering ten risk scenarios. Sustainability performance of turboprop aircraft is detailed in~\cite{Aydin2013}. In~\cite{Barke2020}, alternative green battery technologies are evaluated for their contributions to the Sustainable Development Goals. Lastly,~\cite{Munk2021} presents a topology optimization-based design for load-bearing components, assuming that reducing maximum stress improves load-carrying capacity, fatigue life, and overall sustainability.

From the reviewed literature, it is evident that existing research on aircraft sustainability has primarily concentrated on fuel types and propulsion systems, with most assessments focusing on operational factors rather than structural components. Studies addressing the sustainability of structural parts and mechanical design often operate under the assumption that enhancing strength and reducing weight inherently lead to improved sustainability. Notably, only the work in~\cite{Swastanto2024} offers a holistic sustainability assessment of a structural component by considering a broad range of criteria.

In this study, we propose a comprehensive sustainability assessment methodology specifically for aircraft parts, demonstrated through an application to a fuselage panel. With minor adjustments, this methodology can be applied to any structural component. Key innovations in the present work include:
\begin{itemize}
	\item The methodology is holistic as it efficiently combines environment, cost, and performance.
	\item The methodology is robust as all combinations of MCDM, and weighting methods consistently identify the most sustainable options.
	\item The methodology evaluates diverse assessment criteria using various MCDM methods, particularly the novel R-TOPSIS method and employs five different objective weighting methods to examine their influence on ranking outcomes.
	\item It integrates SimaPro software with the Ecoinvent 3 database, a reliable combination, to conduct both Life-Cycle Analysis (LCA) and Life-Cycle Costing (LCC).
	\item These innovations contribute to a more nuanced and adaptable approach to sustainability assessment in the field of aircraft structural design.
\end{itemize}

\section{The technological problem}\label{sec:technological_problem}
The objective of this work was to develop a robust methodology for assessing the sustainability of aircraft parts. 
While this methodology functions independently, it can also serve as a foundational step toward creating an eco-design methodology for aircraft components.

This work is motivated by the Clean Aviation project FASTER-H2, which aims to validate, downselect, mature, and demonstrate key technologies, as well as provide architectural integration for an ultra-efficient, hydrogen-enabled fuselage and empennage designed for targeted ultra-efficient short-to-medium range (SMR) aircraft. The current study addresses a specific objective within this project: the development of a sustainability-driven eco-design approach for a fuselage that integrates a liquid hydrogen storage tank.

\subsection{Geometries and materials}
\begin{figure}[htbp]
	\centering
	\includegraphics[width=0.8\textwidth]{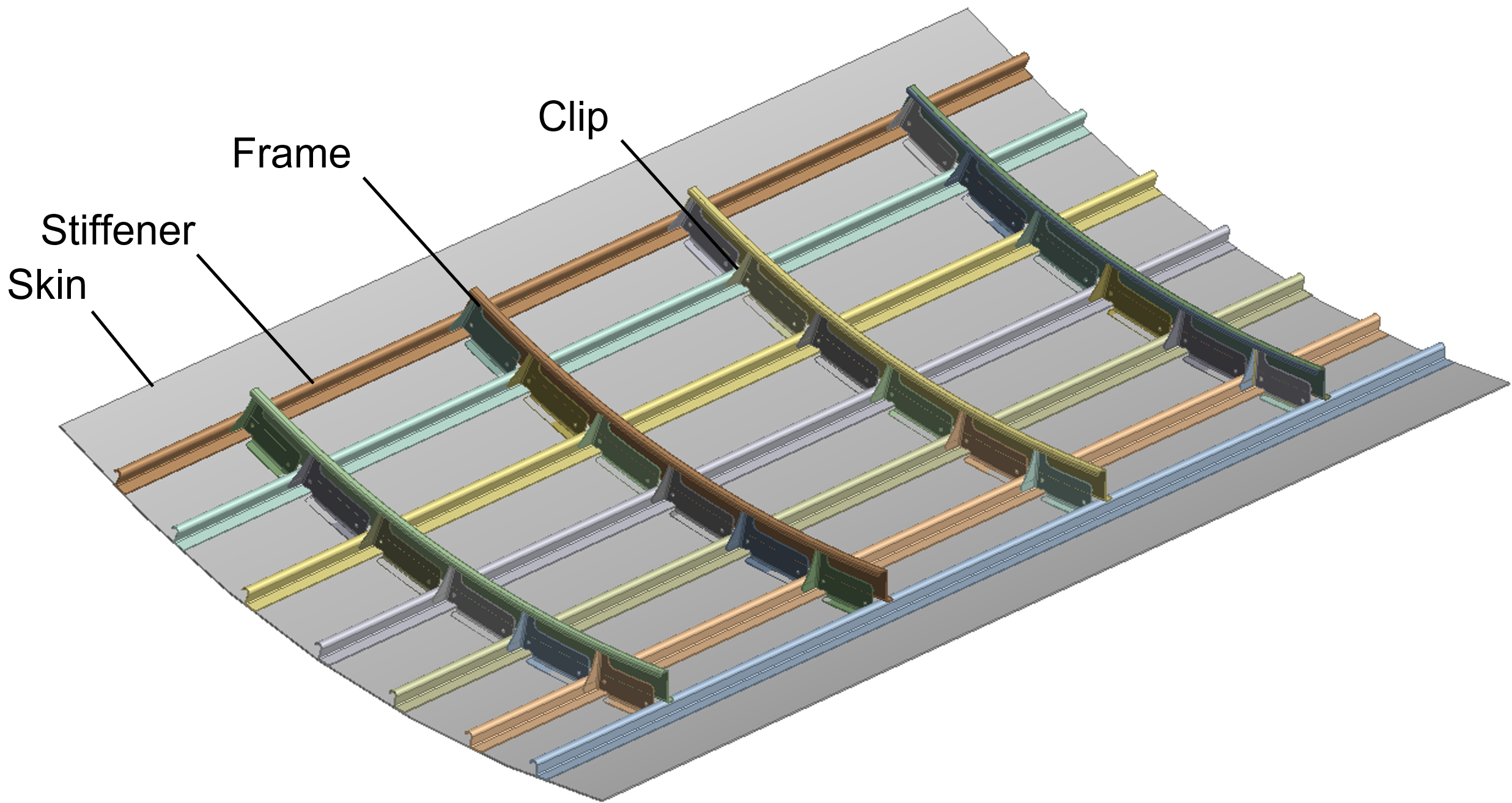}
	\caption{Sketch of the panel}
	\label{fig:panel_sketch}
\end{figure}

The panel is comprised of a curved skin, 7 stiffeners of Z-profile, 4 frames, and 24 clips. A sketch of the panel is depicted in Fig.~\ref{fig:panel_sketch}. The following alternative materials and joining methods manufacturing processes are considered:

\begin{itemize}
	\item Aluminum 2024 (Al2024), manufactured using stretch forming,
	\item Aluminum 7075 (Al7075), manufactured using hydroforming
	\item Thermosetting (TS) composite material, manufactured using autoclave
	\item Thermoplastic (TP) composite material, manufactured using autoclave
	\item Welding for aluminum and TP
	\item Bonding for TS
\end{itemize}

Also, for the skin, the stiffeners, and the frames, three different thickness values were considered. By alternating between the three materials, the three joining techniques, and the three thickness values, nine distinct alternatives were created, as summarized in Table~\ref{tab:panel_alternatives}.

\begin{table}[htbp]
	\caption{Characteristics of the nine panel alternatives}
	\label{tab:panel_alternatives}
	\centering
	\begin{tabular}{l c c c c c c c c}
		\toprule
		Alternative & Skin & Stiffener & Frame & Clip & Joining & Skin & Stiffener & Frame \\
		& Material & Material & Material & Material & Method & Thick. & Thick. & Thick. \\
		& & & & & & (mm) & (mm) & (mm) \\
		\midrule
		S0 & Al2024 & Al7075 & Al2024 & Al2024 & Welding & 2.8 & 2.4 & 1.6 \\
		S1 & Al2024 & Al7075 & Al2024 & Al2024 & Welding & 2.5 & 2.2 & 1.5 \\
		S2 & Al2024 & Al7075 & Al2024 & Al2024 & Welding & 2.8 & 2.4 & 1.6 \\
		\cmidrule(r){1-9}
		S3 & TS & TS & TS & TS & Bonding & 2.8 & 2.4 & 1.6 \\
		S4 & TS & TS & TS & TS & Bonding & 2.5 & 2.2 & 1.5 \\
		S5 & TS & TP & TS & TS & Bonding & 2.8 & 2.4 & 1.5 \\
		\cmidrule(r){1-9}
		S6 & TP & TP & TP & TP & Welding & 2.8 & 2.4 & 1.6 \\
		S7 & TP & TP & TP & TP & Welding & 2.5 & 2.2 & 1.5 \\
		S8 & TP & TP & TP & TP & Welding & 2.8 & 2.4 & 1.6 \\
		\bottomrule
	\end{tabular}
\end{table}

\section{The sustainability assessment methodology}

\subsection{Related work}
The use of Multi-Criteria Decision-Making (MCDM) methods for sustainability assessment in engineering structures is a well-established research area. Ziemba~\cite{Ziemba2022} provided a thorough review that scientifically analyzes the suitability of various Multi-Criteria Decision Analysis (MCDA) methods for sustainability-related decision-making challenges, focusing on sustainability assessment and sustainable development. In engineering, most studies have targeted buildings~\cite{Negrin2023,SanchezGarrido2022} and environmental pollution management frameworks~\cite{Khalili2013}, with relatively few examining mechanical structures~\cite{Filippatos2024,Jasinski2016}. For instance, Negrin et al.~\cite{Negrin2023} utilized unspecified MCDM techniques combined with three distinct weighting methods: equal weights for each objective, a 50-25-25 distribution giving more weight to environmental criteria, and a distribution derived from the CRITIC method~\cite{Diakoulaki1995}, to evaluate the sustainability of different structural designs.

In~\cite{SanchezGarrido2022}, Sanchez-Garrido employed five MCDM methods, introducing a sustainability index created by weighted aggregation of the scores across MCDM methods. The criteria weights were based on their frequency of use in civil engineering MCDM design, as this usage frequency is considered reflective of each method's strengths and limitations, with data gathered from the literature. Bhat et al.~\cite{Bhat2020} applied the TOPSIS MCDM method in conjunction with the Entropy weighting method to evaluate the sustainability of various machining methods. Jasinski et al.~\cite{Jasinski2016} developed a comprehensive sustainability assessment framework for the automotive industry by selecting criteria from the literature and refining them through expert interviews with professionals from academia, automotive manufacturers, consultancies, and NGOs. Their framework aims to serve as a decision-support tool during the early stages of vehicle development. Lastly, Markatos et al.~\cite{Markatos2023} combined the Analytical Hierarchy Method (AHM) MCDM method with a Weighted Sum Model (WSM) for sustainability assessment. These studies illustrate the versatility and adaptability of MCDM methods across various engineering applications, highlighting their potential for supporting sustainability-oriented decision-making.

\subsection{The methodology}
\begin{figure}[htbp]
	\centering
	\includegraphics[width=\textwidth]{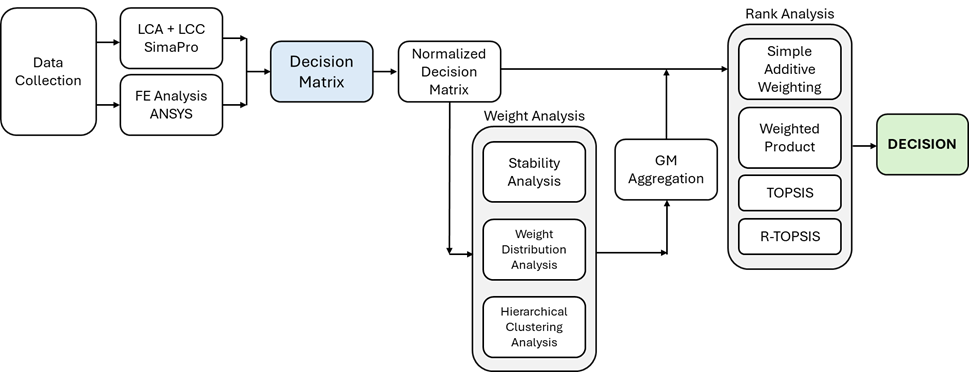}
	\caption{Components and sub-components of the sustainability assessment methodology}
	\label{fig:methodology}
\end{figure}

The flowchart in Figure~\ref{fig:methodology} depicts the basic components and sub-components of the sustainability assessment methodology. In the following sections, all components will be described in detail.

\subsection{Criteria}
While the definition of sustainability is inherently dynamic, it traditionally includes environmental, economic, technical, and social criteria. In the methodology presented here, the focus is on the first three criteria-environmental, economic, and technical—which are detailed in the following sections.

\subsubsection{Environment}
The environmental criteria are grounded in a comprehensive cradle-to-grave Life Cycle Assessment (LCA) performed using SimaPro software with the Ecoinvent 3 library, which is in alignment with ISO standards. The framework for the LCA consists of the following steps:

\paragraph{A. Definition of Goal and Scope} 
The first step of an LCA involves outlining the purpose and scope of the assessment, including the functional unit under study, the system boundaries, and the assumptions and limitations.

\paragraph{B. Inventory Analysis} 
The second step requires the collection of data.

\paragraph{C. Impact assessment} 
In this step are chosen the environmental categories that we want to interpret. This is done by selecting the appropriate method in SimaPro.

\paragraph{D. Interpretation} 
The last step of an LCA is where the results are discussed.

Each of these steps is detailed in the sections that follow.

\paragraph{A. Definition of Goal and Scope} 
The primary objective of this assessment is to evaluate the environmental impact associated with the production of the panel, focusing on variations in materials, geometries, and manufacturing processes for nine panel alternatives. The functional unit in this study is defined as a single panel. The initial design (S0 scenario) of the panel (geometry and materials) was provided in the context of the Faster-H2 project, and the other scenarios were based on the initial one with changes in the geometry and the materials for the scope of the sustainability assessment methodology. This analysis is considered cradle-to-grave and covers all production and manufacturing stages, from raw material extraction until the end of the life of the panel, including its use phase. As use phase is considered the panel's transportation as a part of the fuselage of an A319 aircraft which has a lifetime of 30 years. Details for the use phase and the different waste scenarios for the different materials can be seen in Appendix A in Tables A10 and A11. Transportation impacts are neglected due to data unavailability.

\paragraph{B. Inventory analysis} 
Data collection, the most labor-intensive phase of the LCA, involves gathering comprehensive information on materials and processes within the defined system boundaries. Key data sources include the Ecoinvent 3 database, and relevant literature. The panel consists of the skin, the stiffeners, the frames and the clips. All elements can be made from three different materials: aluminum, thermoset CFRP, and thermoplastic CFRP. Materials and manufacturing processes for each part of the alternatives described in Table~\ref{tab:panel_alternatives} are listed in Tables~\ref{tab:materials_S0_S2} to~\ref{tab:materials_S6_S8}. More data for the materials used in the LCA are provided in Appendix A1.

\begin{table}[h!]
	\caption{Materials and manufacturing methods used for alternatives S0, S1, and S2}
	\label{tab:materials_S0_S2}
	\centering
	\begin{tabular}{lll}
		\toprule
		Part & Material & Manufacturing process \\
		\midrule
		Skin Panel & Al2024 & Stretch Forming \\
		Stringer & Al7075 & Hydroforming \\
		Clip & Al2024 & Incremental Sheet Forming \\
		Frame & Al2024 & Hydroforming \\
		\bottomrule
	\end{tabular}
\end{table}

\begin{table}[h!]
	\caption{Materials and manufacturing methods used for alternatives S3, S4 and S5}
	\label{tab:materials_S3_S5}
	\centering
	\begin{tabular}{lll}
		\toprule
		Part & Material & Manufacturing process \\
		\midrule
		Skin Panel & TS prepreg 57.7\% wt. cf. & Autoclave \\
		Stringer & TS prepreg 57.7\% wt. cf. & Autoclave \\
		Clip & TS prepreg 57.7\% wt. cf. & Autoclave \\
		Frame & TS prepreg 57.7\% wt. cf. & Autoclave \\
		\bottomrule
	\end{tabular}
\end{table}

\begin{table}[h!]
	\caption{Materials and manufacturing methods used for alternatives S6, S7 and S8}
	\label{tab:materials_S6_S8}
	\centering
	\begin{tabular}{lll}
		\toprule
		Part & Material & Manufacturing process \\
		\midrule
		Skin Panel & TP prepreg 57\% wt. cf. & Autoclave \\
		Stringer & TP prepreg 57\% wt. cf. & Autoclave \\
		Clip & TP prepreg 57\% wt. cf. & Autoclave \\
		Frame & TP prepreg 57\% wt. cf. & Autoclave \\
		\bottomrule
	\end{tabular}
\end{table}

\begin{figure}[htbp]
	\centering
	\includegraphics[width=0.9\textwidth]{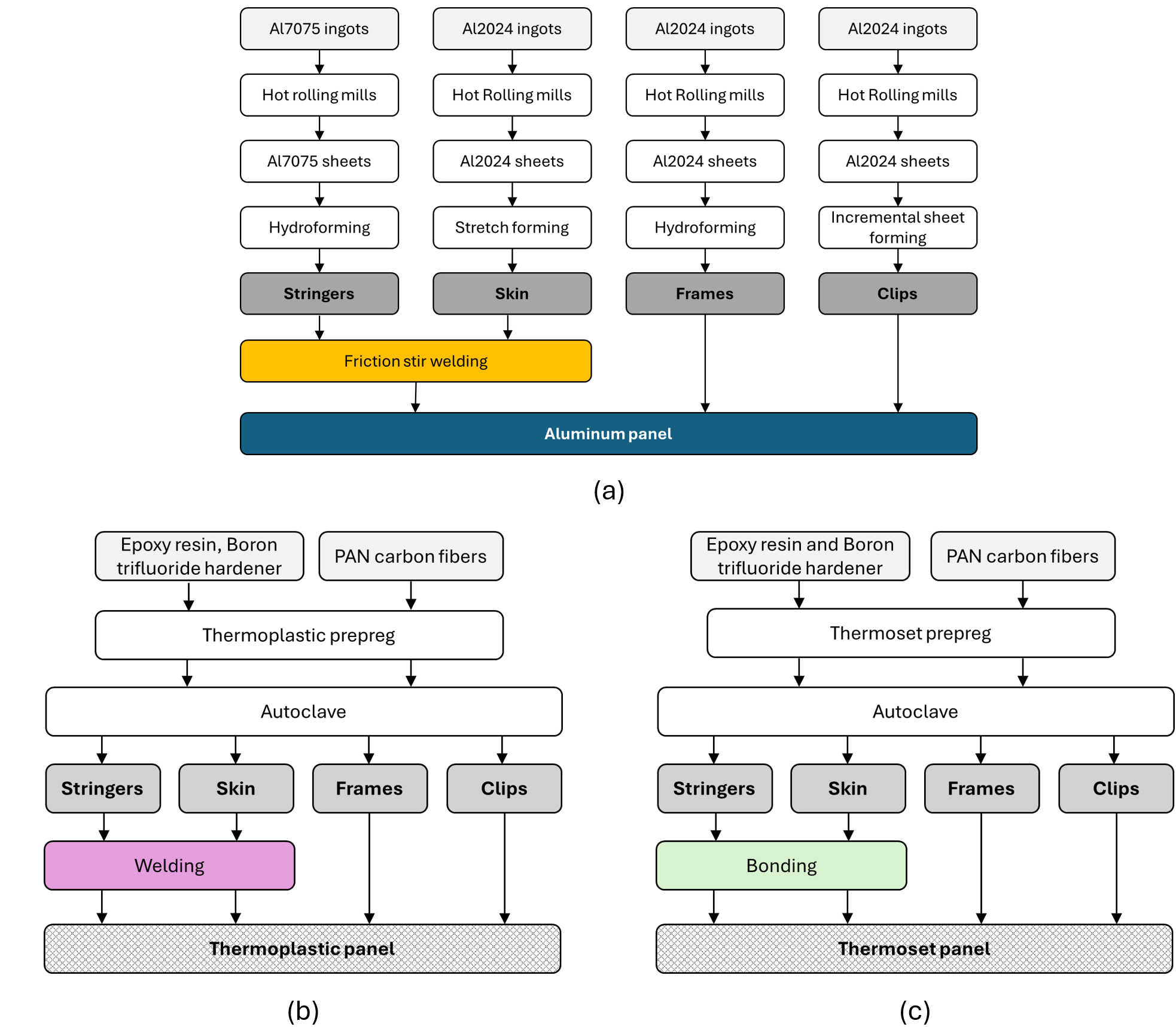}
	\caption{Manufacturing process tree of (a) the aluminum panel and (b) the thermoplastic CFRP panel}
	\label{fig:process_tree}
\end{figure}

LCA also encompasses pre-production processes essential for preparing materials in the required forms for manufacturing. These processes are schematically explained in Figure~\ref{fig:process_tree}. In summary:

For the aluminum panel:
\begin{itemize}
	\item Stiffeners: Al7075 ingots are passed through rolling mills to form sheets, which then are used to produce the stiffeners and clips through the hydroforming process.
	\item Skin: Al2024 ingots are passed through rolling mills to form sheets, which then are used to produce the skin through the stretch forming process.
	\item Frames: Al2024 ingots are passed through rolling mills to form sheets, which then are used to produce the stiffeners and clips through the hydroforming process.
	\item Clips: Al2024 ingots are passed through rolling mills to form sheets, which then are used to produce the stiffeners and clips through the incremental sheet forming process.
\end{itemize}

Stiffeners are welded to the skin through friction stir welding.

For the thermoplastic panel:
\begin{itemize}
	\item Polyphenylene powder, a thermoplastic resin, is added to the PAN carbon fiber fabric and the prepreg is constructed. Then, using autoclave the skin, stiffeners, frames and clips are manufactured. Finally, the stiffeners are welded to the skin.
\end{itemize}

For the thermoset panel:
\begin{itemize}
	\item The process is similar to the process of thermoplastic composite with different joining method. Epoxy resin with Boron trifluoride hardener, is added to the PAN carbon fiber fabric and the prepreg is constructed. Then, using autoclave the skin, stiffeners, frames and clips are manufactured. Finally, the stiffeners are bonded to the skin. Data for the bonding can be found in Table A5.
\end{itemize}

Comprehensive material and process data are provided in the Appendix.

\paragraph{C. Impact assessment}
The environmental criteria considered in this study are:

\begin{itemize}
	\item Human Health (C1): Measured as Disability Adjusted Life Years (DALYs), representing the combined years of life lost and years lived with disability.
	\item Ecosystems (C2): Assessed based on species loss over a specific area and time period.
	\item Resource Scarcity (C3): Evaluated as the additional future production costs of resources over an infinite timeframe, with constant annual production and a 3\% discount rate applied.
	\item Global Warming potential (GWP) (C4): Climate change factors of IPCC method with a timeframe of 100 years, where carbon dioxide uptake is implicitly included.
\end{itemize}

These criteria were calculated using two methods:

ReCiPe 2016 Endpoint (H) V1.08: This method, available in SimaPro, is categorized globally, with characterization factors representing impacts on a global scale. Based on the Hierarchist perspective (H), it aligns with generally accepted policy principles concerning timeframes and related considerations. The ReCiPe method provides results on human health, ecosystems, and resource scarcity.

IPCC 2021 GWP100 V1.02: Falling under SimaPro's Single-Issue category, this method specifically assesses Global Warming Potential (GWP) in terms of kg CO$_2$ equivalent.

\begin{figure}[htbp]
	\centering
	\subfigure[]{
		\includegraphics[width=0.48\textwidth]{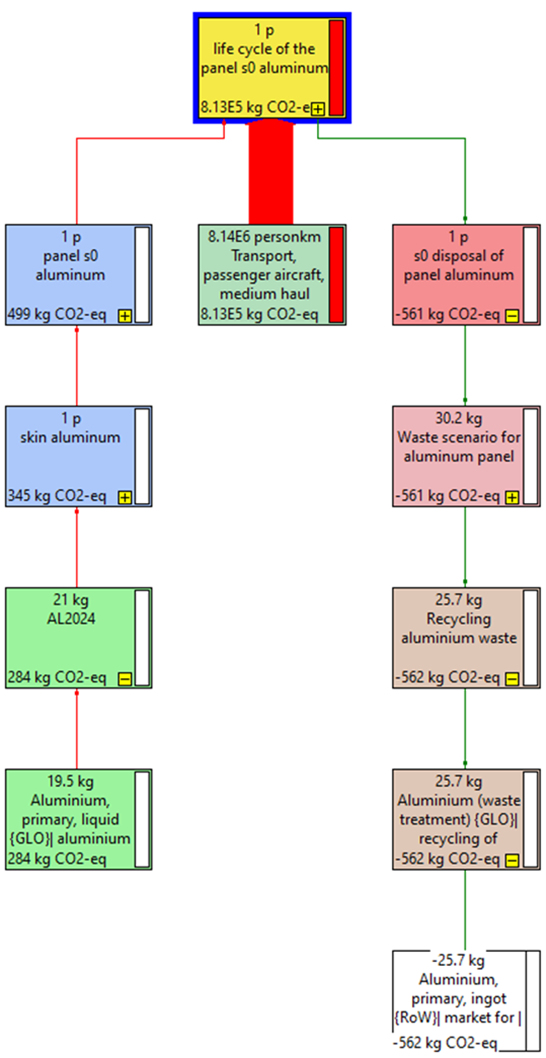}
		\label{fig:simapro_tree_a}
	}
	\subfigure[]{
		\includegraphics[width=0.48\textwidth]{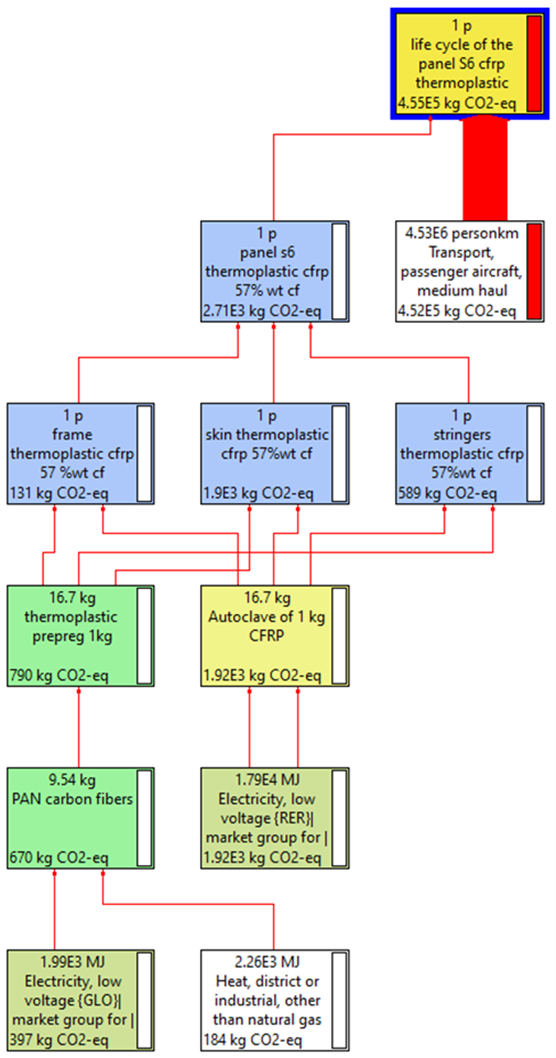}
		\label{fig:simapro_tree_b}
	}
	\caption{SimaPro tree-diagram of the analysis for the computation of the GWP criterion: (a) Aluminum panel, (b) thermoplastic panel. The thickness of the arrows and the bar on the right of the text boxes are analogous to the contribution of each process/material.}
	\label{fig:simapro_tree}
\end{figure}

Fig.~\ref{fig:simapro_tree} illustrates the tree diagram of the SimaPro process used to calculate the GWP criterion for the aluminum panel and thermoplastic panel. The tree diagram for the thermoset panel is omitted here for brevity.

\paragraph{D. Interpretation of the results}
In this study, the interpretation of the results is not straightforward and is derived from ranking the alternative designs of the panel in terms of sustainability, combining environmental, cost, and performance criteria.

\subsubsection{Cost}
LCC is an economic analysis method focused on calculating all costs associated with constructing, operating, and maintaining a project over a specified period~\cite{Markowska2023}. Within this study, an LCC framework is implemented in SimaPro to capture the full range of expenses related to panel production. The cost criteria (C5 to C8) are detailed below:

\begin{itemize}
	\item Material Cost (C5): This includes expenditures on raw materials used in production; see Table A9.
	\item Energy Cost (C6): Accounts for energy consumption throughout the production and manufacturing processes.
	\item Use Cost (C7): Includes kerosene consumption for transporting the panel throughout the entire operational lifetime of the A319.
	\item End of Life (EoL) Cost (C8): Accounts for the cost of EoL services (recycling, landfill, Incineration, etc.).
\end{itemize}

This LCC methodology provides a comprehensive view of economic impacts across the panel's lifecycle, enabling a thorough sustainability assessment. Due to data limitations, labor costs and service costs for manufacturing processes were not included in the study.

\begin{figure}[htbp]
	\centering
	\includegraphics[width=0.8\textwidth]{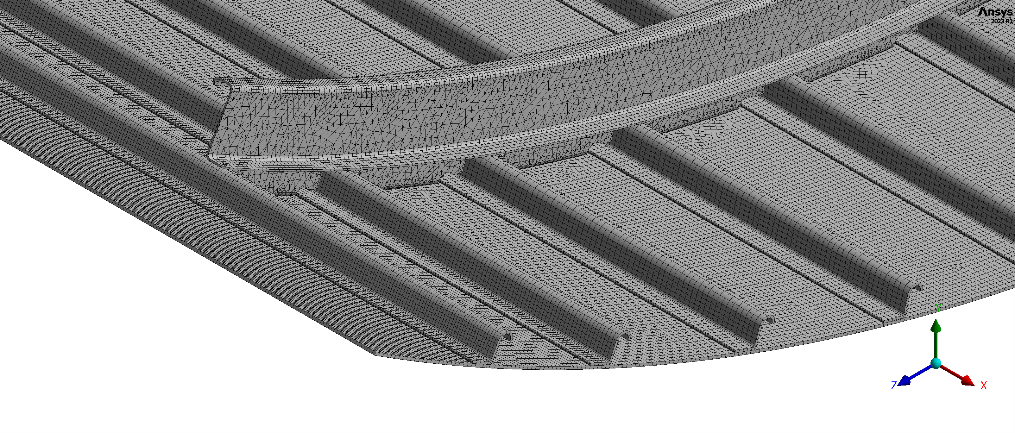}
	\caption{Part of the FE mesh of the panel created using the ANSYS software}
	\label{fig:fem_mesh}
\end{figure}

\begin{figure}[htbp]
	\centering
	\includegraphics[width=\textwidth]{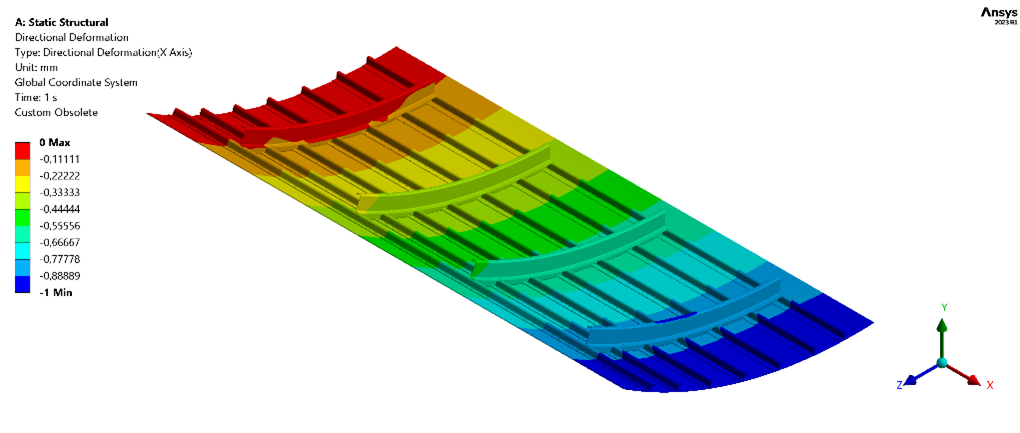}
	\caption{Axial deformation of the panel due to application of the displacement of -1 mm}
	\label{fig:axial_deform}
\end{figure}

\subsubsection{Performance}
The performance of the fuselage panel is evaluated based on its mass and specific stiffness. Mass is designated as performance criterion C9, as it is a critical factor in sustainability assessments. A lighter panel requires fewer materials, less energy, and lower costs, resulting in reduced emissions. Additionally, reducing panel weight enhances aircraft efficiency by decreasing fuel consumption during operation, resulting in less environmental impacts during the use phase. Specific stiffness, an essential structural design parameter, is also evaluated. To compute specific stiffness, a finite element model (FEM) of the panel was developed in ANSYS software, as shown in Fig.~\ref{fig:fem_mesh}, with all components represented by solid elements (excluding clips). Specific stiffness was determined by applying a unit axial compressive displacement at one end of the panel while fully constraining the opposite end. Fig.~\ref{fig:axial_deform} illustrates the axial deformation contour of the panel under this load. Specific stiffness is designated as performance criterion C10.

Table~\ref{tab:criteria_summary} summarizes the 10 criteria in terms of category, impact type and sign.
\begin{table}
	\caption{Criteria description and categorization for the MCDM analysis}
	\label{tab:criteria_summary}
	\centering
	\begin{tabular}{lllr}
		\toprule
		Criteria & Description & Category & Sign \\
		\midrule
		C1 & Human Health (DALY's) & Environment & -1 \\
		C2 & Ecosystems (species.year) & Environment & -1 \\
		C3 & Resources (USD 2013) & Environment & -1 \\
		C4 & Global Warming Potential (kg CO$_2$) & Environment & -1 \\
		C5 & Material Cost & Cost & -1 \\
		C6 & Energy Cost & Cost & -1 \\
		C7 & Use cost Cost & Cost & -1 \\
		C8 & EoL Cost & Cost & -1 \\
		C9 & Mass & Performance & -1 \\
		C10 & Specific Stiffness & Performance & +1 \\
		\bottomrule
	\end{tabular}
\end{table}

\subsection{MCDM analysis}
\begin{figure}[h!]
	\centering
	\includegraphics[width=0.7\textwidth]{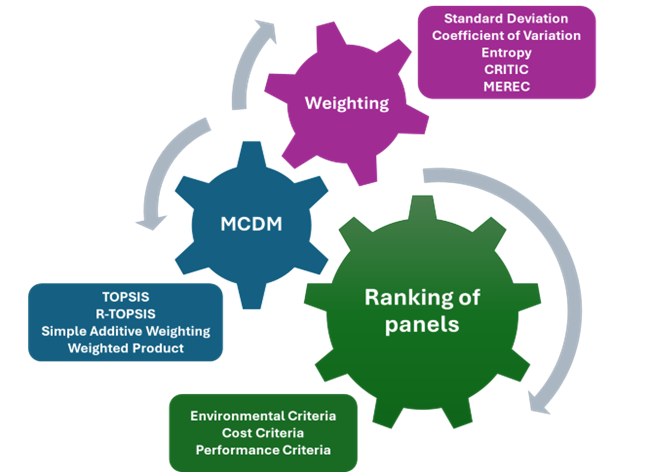}
	\caption{The basic components of MCDM analysis}
	\label{fig:mcdm_components}
\end{figure}

Assessing the sustainability of aircraft components involves multiple criteria, including environmental, economic, and performance metrics. To assess these criteria and make appropriate decisions, we use a set of Multi-Criteria Decision-Making (MCDM) tools. Our approach, schematically explained in Fig.~\ref{fig:mcdm_components}, utilizes i) different normalization techniques, ii) objective weighting methods, and iii) ranking methodologies to ensure a robust and reliable sustainability assessment.

We begin by formulating the decision matrix $\mathbf{X}=[x_{ij}]_{m\times n}$ that collects the performance values of each alternative across all criteria, i.e., $x_{ij}$ refers to the value of the $j$-th criterion for the $i$-th alternative. The $i=1,2,\ldots,m$ denote the set of alternatives, where $m=9$ corresponds to the nine different aircraft panel designs that are being studied, and $j=1,2,\ldots,n$ are the criteria in our case study, $n=10$ includes the sustainability assessment's environmental impacts, costs, and performance metrics.

Because the criteria have different units and scales, normalizing the data is necessary to make meaningful comparisons by transforming the raw data $x_{ij}$ into a dimensionless form $n_{ij}$, ensuring that all criteria contribute appropriately to the decision-making process without being unduly influenced by their original units or magnitudes. The normalized decision matrix $\mathbf{N}=[n_{ij}]_{m\times n}$ is obtained by applying well-established procedures, including vector normalization, min-max normalization, and linear scale normalization. These procedures transform the data into a common scale, fixing the nonuniformity in the performance of different alternatives.

After data normalization, the next step is to determine the weights $w_j$ of the criteria $C_j$ using objective weighting methods in order to capture the relative importance of each criterion. We avoid subjectivity in the weighting by using several methods, which are Standard Deviation (SD), Coefficient of Variation (COV), Entropy, CRITIC, and MEREC, that aim to account for the distribution of variables both across and within the criteria.

Consequently, once the normalized decision matrix $\mathbf{N}$, as well as the weight vector $\mathbf{w}=[w_1,w_2,\ldots,w_n]$ are defined, we apply MCDM ranking methods to evaluate and compare the alternatives. Methods such as Simple Additive Weighting (SAW), Weighted Product (WP), Technique for Order Preference by Similarity to Ideal Solution (TOPSIS), and the novel R-TOPSIS method are used to aggregate the criteria values and produce a ranked list of alternatives based on their overall sustainability performance.

In the subsequent sections, we explain the normalization methods applied (Section 3.3.1), the objective weighting methods used to set the criteria weights (Section 3.3.2), and the MCDM ranking methods employed for the evaluation of the alternatives (Section 3.3.3).

\subsubsection{Normalization methods}
In our sustainability analysis of the design of an aircraft panel, we consider both cost criteria (which should be minimized) and benefit criteria (which should be maximized). Therefore, the normalization methods we employ must effectively address both types of criteria to ensure that the meaning of the normalized values is straightforward, with a higher normalized value indicates better performance across all criteria. Among the many normalization methods available, we have selected three widely recognized approaches: a) vector normalization, b) min-max normalization, and c) linear scale normalization as presented in Table~\ref{tab:normalization_methods}, where $x_{ij}$ denotes the original value, and $n_{ij}$ is the normalized value for the $i$-th alternative with respect to the $j$-th criterion.

\begin{table}[h!]
	\caption{Mathematical formulation of normalization methods for benefit and cost criteria}
	\label{tab:normalization_methods}
	\centering
	\begin{tabular}{llll}
		\toprule
		Method & Benefit criteria (max) & Cost criteria (min) & Description \\
		\midrule
		vector normalization & $n_{ij}=\frac{x_{ij}}{\sqrt{\sum_{i=1}^m x_{ij}^2}}$ & 
		$n_{ij}=1-\frac{x_{ij}}{\sqrt{\sum_{i=1}^m x_{ij}^2}}$ & 
		\begin{tabular}{l}Transforms using\\Euclidean norm,\\with cost criteria\\inverted\end{tabular} \\
		\addlinespace
		linear scale & $n_{ij}=\frac{x_{ij}}{\max_i x_{ij}}$ & 
		$n_{ij}=\frac{\min_i x_{ij}}{x_{ij}}$ & 
		\begin{tabular}{l}Scales relative to\\maximum for\\benefits and\\minimum for costs\end{tabular} \\
		\addlinespace
		min-max & $n_{ij}=\frac{x_{ij}-\min_i x_{ij}}{\max_i x_{ij}-\min_i x_{ij}}$ & 
		$n_{ij}=\frac{\max_i x_{ij}-x_{ij}}{\max_i x_{ij}-\min_i x_{ij}}$ & 
		\begin{tabular}{l}Scales to [0,1] range\\with appropriate\\direction\end{tabular} \\
		\bottomrule
	\end{tabular}
\end{table}

In our case study, the criteria are classified as follows:
\begin{itemize}
	\item Benefit criterion (to be maximized): $C_{10}$ (Specific stiffness)
	\item Cost criteria (to be minimized): $C_1-C_9$ Environmental impacts (Human Health, Ecosystems, Resources, Global Warming Potential), costs (Material, Energy, Use-phase, End-of-life), and mass efficiency.
\end{itemize}

The normalized decision matrix $\mathbf{N}=[n_{ij}]$, that is produced from all three methods, as detailed above, contains dimensionless values comparable across all criteria, with higher values consistently indicating better performance. However, in MCDM analyses, it is important to select the appropriate normalization method which can greatly affect both the final ranking of alternatives and the calculation of criteria weights using objective weighting methods. Previous studies have demonstrated the fact that in MCDM applications, different normalization methods produce relatively different results, which then affect the robustness and reliability when making a decision~\cite{Jahan2015,Milani2005}.

To identify the best-applied normalization technique in our study, we performed an extensive stability analysis, presented in Section 4.2.1; to evaluate how sensitive the weighting methods and overall rankings are to perturbations in the normalized data. Our simulations indicated that the choice of normalization method significantly impacts the performance of objective methods used for calculating weights, especially for those methods based on the data's spread. Applying incorrect normalization may result in biased or unstable weights, which can affect the reliability of the MCDM process~\cite{Zanakis1998}.

For our dataset, vector normalization proved the best method since it showed more stability in generating criteria weights across various perturbation levels compared to min-max and linear scale normalization. Moreover, we verified that all MCDM methods that are applied for ranking the alternatives must utilize the same normalization method used through the weighting stage. Based on these findings, we adopted vector normalization as the method for normalization in this study.

\subsubsection{Objective weighting methods}
Determining the weights is one of the most crucial steps because it affects the ranking of the alternatives and can be determined subjectively or objectively. Subjective weights are based on the opinion of experts, whereas objective weights are based on the data properties like variability or contrast. In this study, we used five well-known objective methods to minimize bias and, therefore, transparency in the whole process of the sustainability assessment. Specifically, the applied methods are standard deviation (SD), coefficient of variation (COV), entropy, CRITIC, and MEREC. These methods were implemented using custom MATLAB functions developed for this project. In the following, we present a brief description of each implemented method.

\paragraph{Standard deviation (SD)}~\cite{Diakoulaki1995,Zeleny1982,Wang2010}
Let $n_{ij}$ be the normalized values, then the weight $w_j$ for criterion $j$ is calculated as
\begin{equation}
	w_j=\frac{s_j}{\sum_{k=1}^n s_k}
\end{equation}
where $s_j=\sqrt{\frac{1}{m-1}\sum_{i=1}^m(n_{ij}-\bar{n}_j)^2}$ is the sample standard deviation of the $j$-th criterion and $\bar{n}_j=\frac{1}{m}\sum_{i=1}^m n_{ij}$ is the mean of the normalized values. The rationale of the SD method is to assign higher weights to criteria with greater variability across alternatives, providing, in this way, a discriminating power that is particularly effective when the differentiation between alternatives is a key consideration.

\paragraph{Coefficient of variation (COV)}~\cite{Zeleny1982}
The COV method extends the SD by considering the relative variability of each criterion since it normalizes the standard deviation with the mean. The weight $w_j$ for $j$-the criterion is given by
\begin{equation}
	w_j=\frac{CV_j}{\sum_{k=1}^n CV_k}
\end{equation}
where $CV_j=\frac{s_j}{\bar{n}_j}$ is the coefficient of variation for the criterion $j$ with $s_j$ and $\bar{n}_j$ defined as in the SD method. Consequently, any criterion where COV exhibits greater values than others is more significant, as its relative weight within the COV is higher, resulting in the enhanced discriminating power of the criterion in differentiating the alternatives.

\paragraph{Entropy method}~\cite{Zeleny1982,Hwang1981}
The Entropy method measures the information content in each criterion using Shannon's entropy concept, where the key idea is to assign weights based on the amount of information present in the criteria. To this end, first, we calculate the proportion $p_{ij}$ of alternative $i$ for criterion $j$
\begin{equation}
	p_{ij}=\frac{n_{ij}}{\sum_{i=1}^m n_{ij}}, \quad \forall i,j
\end{equation}
and then compute the entropy $e_j$ of each criterion:
\begin{equation}
	e_j=-k\sum_{i=1}^m p_{ij}\ln(p_{ij}), \quad \forall j
\end{equation}
where the constant $k=\frac{1}{\ln(m)}$ guarantees that $0\leq e_j \leq 1$. The degree of diversification $d_j$ of the information contained in each criterion is $d_j=1-e_j$, and finally, the normalized weight $w_j$ is calculated by
\begin{equation}
	w_j=\frac{d_j}{\sum_{k=1}^n d_k}
\end{equation}

Generally, the lower the information entropy of a criterion, the more information it provides, so the greater its weight, while criteria with a more uniform distribution of values (higher entropy) are considered less informative and assigned lower weights.

\paragraph{CRITIC method}~\cite{Diakoulaki1995}
The CRITIC (CRiteria Importance Through Intercriteria Correlation) method calculates objective weights for criteria by considering both standard deviation and inter-criteria correlations, by aiming to reflect the significance of individual criteria as well as the conflicts that may arise between them. To this end, the weight for each criterion is computed using these two factors. For each criterion $j$, the weight $w_j$ is calculated as:
\begin{equation}
	w_j=\frac{C_j}{\sum_{k=1}^n C_k}
\end{equation}
where $C_j$ is the amount of information contained in $j$-th criterion is determined by:
\begin{equation}
	C_j=s_j\sum_{k=1}^n(1-r_{jk})
\end{equation}

Here, $s_j$ is the sample standard deviation of $j$-th criterion as defined in the SD method, and $r_{jk}$ represents the Pearson correlation coefficient between criterion $j$ and $k$ computed as:
\begin{equation}
	r_{jk}=\frac{\sum_{i=1}^m(n_{ij}-\bar{n}_j)(n_{ik}-\bar{n}_k)}{\sqrt{\sum_{i=1}^m(n_{ij}-\bar{n}_j)^2\sum_{i=1}^m(n_{ik}-\bar{n}_k)^2}}
\end{equation}

CRITIC assigns higher weights to criteria with high standard deviation (which measures contrast intensity) and low correlation with other criteria (which measure conflict). The factor $(1-r_{jk})$ emphasizes criteria that are less correlated (lower $r_{jk}$) with others, hence increasing their weights to reflect their distinct contribution to the decision. These two facts make CRITIC particularly effective for problems where both the discriminating power of criteria and their independence from other criteria are crucial to avoid information redundancy. However, it is necessary to note that CRITIC mainly identifies a linear relationship in the pairs of variables through correlations, making it less effective in situations that involve multiple criteria with complex interdependencies in a non-linear way.

\paragraph{MEREC method}~\cite{KeshavarzGhorabaee2021,KeshavarzGhorabaee2021b}
The MEREC (MEthod based on the Removal Effects of Criteria) is a novel objective weighting technique that determines criteria weights by considering the effect of excluding each criterion on the overall performance of alternatives. MEREC employs a causality-based approach, emphasizing the actual contribution of each criterion to the decision process, in contrast to traditional objective weighting approaches that depend exclusively on direct variability measurements like standard deviation or entropy.

For each alternative $i$, the overall performance score $S_i$ is defined and computed via a nonlinear function $f(x)=\ln(1+|\ln(x)|)$ as follows:
\begin{equation}
	S_i=\ln\left(1+\frac{1}{m}\sum_{j=1}^m|\ln(n_{ij})|\right)
\end{equation}
where $n_{ij}$ are the normalized values. The performance score of the alternative $i$ when $j$th criterion is removed is calculated as:
\begin{equation}
	S'_{ij}=\ln\left(1+\frac{1}{m}\sum_{\substack{k=1\\k\neq j}}^m|\ln(n_{ik})|\right)
\end{equation}

Then, the effect of removing the $j$th criterion is measured as the summation of absolute deviations $E_j=\sum_{i=1}^n|S'_{ij}-S_i|$ and the weight $w_j$ is determined by normalizing the removal effects:
\begin{equation}
	w_j=\frac{E_j}{\sum_{k=1}^m E_k}
\end{equation}

Obviously, higher removal effects $E_j$ indicate greater importance of a criterion in differentiating alternatives, leading to higher weights. The method utilizes a logarithmic metric to aggregate performances and assesses the significance of each criterion by evaluating its impact on decision outcomes when removed. The results of the simulation that the authors carried out in~\cite{KeshavarzGhorabaee2021} show that MEREC produces reliable weights that exhibit a strong correlation with other (SD, Entropy, CRITIC) for small problems; however, this correlation weakens as the problem size increases. The effectiveness of any weighting method depends on the proper normalization of the decision matrix and the careful execution of the computational procedure.

\subsubsection{MCDM methods}
After obtaining the normalized decision matrix $\mathbf{N}=[n_{ij}]_{m\times n}$ and criteria weights $\mathbf{w}=[w_1,w_2,\ldots,w_n]$, four MCDM methods were employed in this study to rank the alternatives: Simple Additive Weighting (SAW), Weighted Product (WP), TOPSIS (Technique for Order of Preference by Similarity to Ideal Solution), and R-TOPSIS (Robust TOPSIS). Each evaluation method handles cost and benefit weighting through different aggregation and ranking mechanisms. In the SAW method, linear aggregation is applied by adding the products obtained from each criterion weighting to a set of normalized measures. On the other hand, the WP method employs a multiplicative aggregation by taking the product of the weighted normalized criteria rather than simply summing them. In contrast, TOPSIS and its strong variant R-TOPSIS concentrate on how far the given alternatives are from the ideal and the anti-ideal alternatives, respectively, and thus evaluate how close they are to these two solutions.

As noted by several researchers~\cite{Zanakis1998,Triantaphyllou1996} different MCDM may provide an inconsistent order of preferences even for the same problem. It is natural since each method has its own aggregator structure which is responsible for combining the information with the criteria. Therefore, applying more than one approach strengthens the reliability of the ranking results by thoroughly studying the problem for the decision using various approaches. To do so, all methods have been applied in a consistent manner by normalizing the decision matrix with respect to the benefit and cost criteria as described in Section 3.3.1. In the following, we present a mathematical description of each method.

\paragraph{Simple additive weighting (SAW)}~\cite{Hwang1981,Yeh2002}
The SAW method is one of the oldest techniques in the MCDM field, and its popularity is owing to its relative simplicity and ease of computation. For each alternative $i$, an overall preference score $V_i$ is calculated by multiplying each normalized value $n_{ij}$ with its associated weight criterion $w_j$ and summing across all criteria:
\begin{equation}
	V_i=\sum_{j=1}^n w_j n_{ij}, \quad i=1,2,\ldots,m
\end{equation}

The alternatives are ranked based on their resulting values $V_i$ in descending order, where a higher value indicates better performance. The SAW method validity is based on the additive utility assumption~\cite{Triantaphyllou2000}; although in many engineering problems, these assumptions do not hold—such as when criteria are interdependent or exhibit non-linear relationships-the SAW method remains a widely used approach in MCDM due to its simplicity and ease of interpretation.

\paragraph{Weighted product (WP)}~\cite{Yeh2002,Triantaphyllou2000}
The WP method is similar to SAW but aids in extending it by means of employing a multiplicative aggregation approach, which effectively captures non-linear relationships between criteria. For each alternative $i$, the preference score $P_i$ is computed as a product of every normalized criterion value raised to the weight corresponding to that criterion:
\begin{equation}
	P_i=\prod_{j=1}^n(n_{ij})^{w_j}, \quad i=1,2,\ldots,m
\end{equation}
where $n_{ij}$ is the normalized value of the $i$th alternative with respect to the $j$th criteria, and $w_j$ is the corresponding criterion weight, ensuring that $\sum_{j=1}^n w_j=1$. In order to make all the alternatives comparable, preference scores $P_i$ are normalized by dividing each $P_i$ by the maximum score within the set:
\begin{equation}
	V_i=\frac{P_i}{\max_{k=1}^m P_k}, \quad i=1,2,\ldots,m
\end{equation}

This normalization guarantees that all the $V_i$ scores fall between 0 and 1, where a higher score represents a better performance; therefore, alternatives are then ranked based on their $V_i$ values in descending order. The WP has two main advantages; first, it penalizes alternatives that perform poorly with higher penalties than the additive technique; second, its geometric aggregation approach makes it ideal for problems where criteria depend on each other~\cite{Triantaphyllou2000}.

\paragraph{TOPSIS}~\cite{Hwang1981,Triantaphyllou2000}
The TOPSIS method is a widely utilized MCDM method that ranks alternatives based on relative closeness to a positive ideal solution (PIS) and a negative ideal solution (NIS). TOPSIS is carried out through the following for a normalized decision matrix $\mathbf{N}=[n_{ij}]_{m\times n}$:

First, calculate the weighted normalized matrix $\mathbf{Y}=[y_{ij}]_{m\times n}$ by applying the criteria weights $w_j$ to each normalized value:
\begin{equation}
	y_{ij}=w_j\cdot n_{ij}, \text{ for } i=1,2,\ldots,m; j=1,2,\ldots,n
\end{equation}

The positive ideal solution $(A^+)$ and negative ideal solution $(A^-)$ are then determined by:
\begin{equation}
	A_j^+=\begin{cases}
		\max_{i=1}^m y_{ij} & \text{if } j \text{ is a benefit criterion} \\
		\min_{i=1} y_{ij} & \text{if } j \text{ is a cost criterion}
	\end{cases};
	A_j^-=\begin{cases}
		\min_{i=1}^m y_{ij} & \text{if } j \text{ is a benefit criterion} \\
		\max_{i=1} y_{ij} & \text{if } j \text{ is a cost criterion}
	\end{cases}
\end{equation}

The separation between each alternative is measured by the Euclidean distances and calculated as
\begin{equation}
	S_i^+=\sqrt{\sum_{j=1}^n(A_j^+-y_{ij})^2}, \quad S_i^-=\sqrt{\sum_{j=1}^n(y_{ij}-A_j^-)^2}, i=1,2,\ldots,m
\end{equation}
to quantify how close each alternative is to PIS and NIS. Finally, the relative closeness coefficient $C_i$ of each alternative to the ideal solution is given by
\begin{equation}
	C_i=\frac{S_i^-}{S_i^++S_i^-}, \quad i=1,2,\ldots,m
\end{equation}

A higher $C_i$ indicates that the alternative $A_i$ is closer to the PIS and further from the NIS, thus being more desirable; therefore, the alternatives are ranked in descending order of $C_i$ values.

\paragraph{R-TOPSIS}~\cite{Aires2019}
R-TOPSIS addresses the rank reversal problem through a domain-based normalization while maintaining the core principles of classical TOPSIS. The method requires a predefined domain $D=[d_{1j},d_{2j}]$ for each criterion $j$ and performs Max or Max-Min normalization across all criteria, with criterion type differentiation (cost or benefit) occurring only in determining ideal solutions (PIS and NIS). The key difference that characterizes both benefit and cost criteria is only about the ideal solutions PIS and NIS since these ideal points remain fixed (see Step 4).

Here are the main steps of the R-TOPSIS algorithm~\cite{Aires2019}:

\textbf{Step 1:} Define the decision matrix $\mathbf{X}=[x_{ij}]_{m\times n}$; the criteria weights as $\mathbf{W}=[w_j]_{1\times n}$, where $w_j>0$ and $\sum_{j=1}^n w_j=1$; and a sub-domain of real numbers $D=[d_j]_{2\times n}$, where $d_j\in\mathbb{R}$, to evaluate the rating of the alternatives, where $d_{1j}$ is the minimum of $D_j$ and $d_{2j}$ is the maximum of $D_j$.

\textbf{Step 2:} Calculate the normalized decision matrix $(n_{ij})$ by using Max or Max-Min as:
\begin{equation}
	n_{ij}=\frac{x_{ij}}{d_{2j}}, \text{ or } n_{ij}=\frac{x_{ij}-d_{1j}}{d_{2j}-d_{1j}}, \quad i=1,2,\ldots,m; j=1,2,\ldots,n
\end{equation}

\textbf{Step 3:} Calculate the weighted normalized decision matrix as
\begin{equation}
	y_{ij}=w_j\cdot n_{ij}, \text{ for } i=1,2,\ldots,m; j=1,2,\ldots,n
\end{equation}

\textbf{Step 4:} Set the negative (NIS) and positive (PIS) ideal solutions such as:
\begin{equation}
	A_j^+=\begin{cases}
		w_j & \text{if } j \text{ is a benefit criterion} \\
		\left(\frac{d_{1j}}{d_{2j}}\right)w_j & \text{if } j \text{ is a cost criterion}
	\end{cases};
	A_j^-=\begin{cases}
		\left(\frac{d_{1j}}{d_{2j}}\right)w_j & \text{if } j \text{ is a benefit criterion} \\
		w_j & \text{if } j \text{ is a cost criterion}
	\end{cases}
\end{equation}

\textbf{Step 5:} Calculate the distances of each alternative $i$ in relation to the ideal solutions such as:
\begin{equation}
	S_i^+=\sqrt{\sum_{j=1}^n(A_j^+-y_{ij})^2}, \quad S_i^-=\sqrt{\sum_{j=1}^n(y_{ij}-A_j^-)^2}, i=1,2,\ldots,m
\end{equation}

\textbf{Step 6:} Derive the closest coefficient of the alternatives $(C_i)$ as
\begin{equation}
	C_i=\frac{S_i^-}{S_i^++S_i^-}, \quad i=1,2,\ldots,m
\end{equation}

Sort the alternatives in descending order. The highest $C_i$ gives the best performance with regard to the evaluation criteria.


\section{Results}

\subsection{Construction and normalization of the decision matrix}
According to the methodology outlined in Section 3.2, the initial decision matrix was constructed using the criteria defined in Table~\ref{tab:criteria_summary}, which consisted of ten criteria (C1-C10) divided into three categories: environmental impact, cost, and performance. Nine alternatives were evaluated, a reference aluminum panel (S0), two aluminum variants (S1-S2), three thermoset CFRP panels (S3-S5), and three thermoplastic CFRP panels (S6-S8). Criteria C1 to C4 focus on environmental factors, emphasizing their importance in ecological evaluations, whereas C5 to C8 address economic issues, reflecting financial considerations. Criterion C9 is the panel's mass, which must also be minimized, while criterion C10 represents specific stiffness, which should be maximized. These latter criteria are classified as performance indicators.

The decision matrix incorporates environmental impact indicators derived from life cycle assessment (C1-C4), comprehensive cost factors covering the entire lifecycle (C5-C8), and key performance parameters (C9-C10). As shown in Table~\ref{tab:initial_matrix}, the data spans different orders of magnitude and units of measurement, necessitating a normalization procedure for meaningful comparison. For instance, the environmental indicators range from very small values (C2, in the order of $10^{-3}$) to very large values (C3 and C4, in the order of $10^5$).

\begin{table}[htb]
	\caption{Initial decision matrix for panel alternatives}
	\label{tab:initial_matrix}
	\centering
	\begin{tabular}{lllllllllll}
		\toprule
		Alternative & C1 & C2 & C3 & C4 & C5 & C6 & C7 & C8 & C9 & C10 \\
		\midrule
		S0 & 1.3300 & 3.430e-3 & 1.160e+5 & 8.130e+5 & 164 & 38.30 & 1.340e+5 & 27.0 & 30.2308 & 4931.845 \\
		S1 & 1.2000 & 3.100e-3 & 1.050e+5 & 7.350e+5 & 153 & 35.90 & 1.210e+5 & 25.2 & 27.3357 & 4897.502 \\
		S2 & 1.3000 & 3.360e-3 & 1.140e+5 & 7.950e+5 & 161 & 36.80 & 1.310e+5 & 26.4 & 29.5650 & 5037.188 \\
		S3 & 0.7560 & 1.950e-3 & 6.560e+4 & 4.610e+5 & 1540 & 1490 & 7.540e+4 & 1.02 & 17.0330 & 7355.105 \\
		S4 & 0.6840 & 1.760e-3 & 5.930e+4 & 4.170e+5 & 1430 & 1350 & 6.820e+4 & 0.924 & 15.4010 & 7306.868 \\
		S5 & 0.7560 & 1.950e-3 & 6.560e+4 & 4.610e+5 & 1520 & 1460 & 7.380e+4 & 0.999 & 16.6570 & 7515.834 \\
		S6 & 0.7460 & 1.920e-3 & 6.480e+4 & 4.550e+5 & 430 & 1440 & 7.450e+4 & 1.010 & 16.8160 & 6999.713 \\
		S7 & 0.6750 & 1.740e-3 & 5.860e+4 & 4.110e+5 & 389 & 1310 & 6.730e+4 & 0.912 & 15.2050 & 6952.876 \\
		S8 & 0.7300 & 1.880e-3 & 6.340e+4 & 4.450e+5 & 421 & 1410 & 7.280e+4 & 0.987 & 16.4450 & 7152.875 \\
		\bottomrule
	\end{tabular}
\end{table}

Considering the diversity of the criteria and their distinct units of the measurement system, the vector normalization approach was utilized to transform the initial decision matrix into comparative dimensionless units, a choice that will be validated through detailed stability analysis in Section 4.2.1. In a decision matrix $\mathbf{X}=[x_{ij}]_{\text{mxn}}$, with $m$ alternatives and $n$ criteria, each normalized value $n_{ij}$ is calculated using the formula:
\begin{equation}
	n_{ij}=\begin{cases}
		\frac{x_{ij}}{\sqrt{\sum_{i=1}^m x_{ij}^2}} & \text{if } j \text{ is a benefit criterion} \\
		1-\frac{x_{ij}}{\sqrt{\sum_{i=1}^m x_{ij}^2}} & \text{if } j \text{ is a cost criterion}
	\end{cases}
\end{equation}
where $x_{ij}$ is the original value for the $i$-th alternative to the $j$-th criterion.

Table~\ref{tab:normalized_matrix} provides the resulting normalized decision matrix, with values rounded to 4 decimal places for consistency. The normalizing procedure allows direct comparison among all criteria while maintaining the relative proportions of alternatives for each criterion, transforming the varied scales of the original criteria into uniform units.

\begin{table}[htb]
	\caption{Normalized decision matrix using the vector normalization technique}
	\label{tab:normalized_matrix}
	\centering
	\begin{tabular}{lllllllllll}
		\toprule
		Alternative & C1 & C2 & C3 & C4 & C5 & C6 & C7 & C8 & C9 & C10 \\
		\midrule
		S0 & 0.5314 & 0.5316 & 0.5313 & 0.5311 & 0.9394 & 0.9889 & 0.5289 & 0.4061 & 0.5293 & 0.2510 \\
		S1 & 0.5772 & 0.5766 & 0.5758 & 0.5761 & 0.9434 & 0.9896 & 0.5746 & 0.4457 & 0.5744 & 0.2492 \\
		S2 & 0.5420 & 0.5411 & 0.5394 & 0.5415 & 0.9405 & 0.9894 & 0.5395 & 0.4193 & 0.5396 & 0.2563 \\
		S3 & 0.7336 & 0.7337 & 0.7349 & 0.7341 & 0.4307 & 0.5691 & 0.7349 & 0.9776 & 0.7348 & 0.3743 \\
		S4 & 0.7590 & 0.7596 & 0.7604 & 0.7595 & 0.4713 & 0.6096 & 0.7603 & 0.9797 & 0.7602 & 0.3718 \\
		S5 & 0.7336 & 0.7337 & 0.7349 & 0.7341 & 0.4381 & 0.5778 & 0.7406 & 0.9780 & 0.7406 & \textbf{0.3825} \\
		S6 & 0.7372 & 0.7378 & 0.7382 & 0.7376 & 0.8410 & 0.5836 & 0.7381 & 0.9778 & 0.7382 & 0.3562 \\
		S7 & \textbf{0.7622} & \textbf{0.7624} & \textbf{0.7632} & \textbf{0.7630} & 0.8562 & 0.6211 & \textbf{0.7634} & \textbf{0.9799} & \textbf{0.7632} & 0.3538 \\
		S8 & 0.7428 & 0.7433 & 0.7438 & 0.7433 & 0.8444 & 0.5922 & 0.7441 & 0.9783 & 0.7439 & 0.3640 \\
		\bottomrule
	\end{tabular}
\end{table}

Examining the normalized values in Table~\ref{tab:normalized_matrix} resulted in the following observations concerning the established ideal points within each criteria category:

Environmental criteria (C1-C4) reveal a distinct separation between aluminum (S0-S2) and composite (S3-S8) alternatives. The aluminum panels exhibit lower normalized values (0.53-0.57 range) across all environmental indicators, whereas CFRP alternatives show superior ecological performance with higher normalized values (0.73-0.76 range). Panel S7 (CFRP Thermoplastic) demonstrates superior environmental performance, with normalized values consistently exceeding 0.76.

Cost criteria (C5-C8) indicate different trends among various cost components. In terms of material and energy costs (C5-C6), aluminum alternatives (S0-S2) exhibit superior performance, with normalized values exceeding 0.94, whereas CFRP thermoset panels (S3-S5) present lower values, ranging from 0.43 to 0.47. CFRP thermoplastic alternatives (S6-S8) exhibit intermediate material cost performance, ranging from 0.84 to 0.86. In terms of end-of-life costs (C8), a contrary trend is noted, with CFRP alternatives achieving significantly greater normalized values (>0.97) compared to aluminum panels, which range from 0.40 to 0.45.

\subsection{Weighting methods analysis and selection}
To assess the alternatives regarding sustainability, appropriate weights must be assigned to the criteria to reflect their relative importance in the decision-making process. Five objective weighting methods were considered: SD, COV, Entropy, CRITIC, and MEREC. However, before applying any of these methods, testing them for reliability and stability within realistic data perturbations, e.g., 1\%--25\% is important. This additional verification allows us to select the most robust methods for the final sustainability analysis in the subsequent sections of this paper.

The selection process comprises three distinct steps to assess the weighting methods:
\begin{itemize}
	\item First, a stability analysis is conducted through perturbation experiments to determine how each weighting method responds to small variations in the normalized decision matrix on each weighting method selected.
	\item Second, the weight distributions from all methods are analyzed to understand how they allocate relative significance to the different criteria.
	\item Finally, the methods are clustered according to their stability by employing hierarchical clustering based on the derived stability scores. This step provides additional statistical support for methods selection.
\end{itemize}

\subsubsection{Stability analysis}
In this step, an extensive stability analysis was carried out to check the stability of the objective weighting methods and choose the most appropriate normalization strategy by examining three well-established normalization approaches in the literature, as presented in Table~\ref{tab:normalization_methods}.

Our analysis utilized perturbations within the range $\varepsilon \in [0.01,0.25]$ with a step size of 0.01, resulting in 25 unique perturbation levels, and at each level, 1000 Monte Carlo iterations were conducted, yielding a total of 25,000 experiments to achieve statistical significance. For each objective method $M$ at perturbation level $\varepsilon$, the stability score $S^{(M,\varepsilon)}$ was calculated as $S^{(M,\varepsilon)}=1/(1+\mu_{\delta})$, where $\mu_{\delta}$ denotes the mean relative weight change across all criteria for the method $M$ at perturbation level $\varepsilon$.

\begin{table}
	\caption{Summary statistics of stability scores under vector normalization}
	\label{tab:stability_scores}
	\centering
	\begin{tabular}{lllllll}
		\toprule
		Method & mean & median & std & variance & min & max \\
		\midrule
		SD & 0.9486 & 0.9479 & 0.0282 & 7.9343$\times$10$^{-4}$ & 0.9040 & 0.9958 \\
		COV & 0.9495 & 0.9488 & 0.0276 & 7.6423$\times$10$^{-4}$ & 0.9059 & 0.9959 \\
		Entropy & 0.9026 & 0.9006 & 0.0522 & 2.7242$\times$10$^{-3}$ & 0.8213 & 0.9917 \\
		CRITIC & 0.9083 & 0.9157 & 0.0633 & 4.0121$\times$10$^{-3}$ & 0.7956 & 0.9956 \\
		MEREC & 0.9948 & 0.9950 & 0.0032 & 9.9518$\times$10$^{-6}$ & 0.9892 & 0.9996 \\
		\bottomrule
	\end{tabular}
\end{table}

The comparative analysis from the results indicated that vector normalization attained the highest overall stability score (mean = 0.94076, var = 7.1000$\times$10$^{-4}$), outperforming both linear scale transformation (mean = 0.91478, var = 3.6200$\times$10$^{-3}$) and min-max normalization (mean = 0.93376, by excluding the undefined results for MEREC).

\begin{figure}[htbp]
	\centering
	\includegraphics[width=0.8\textwidth]{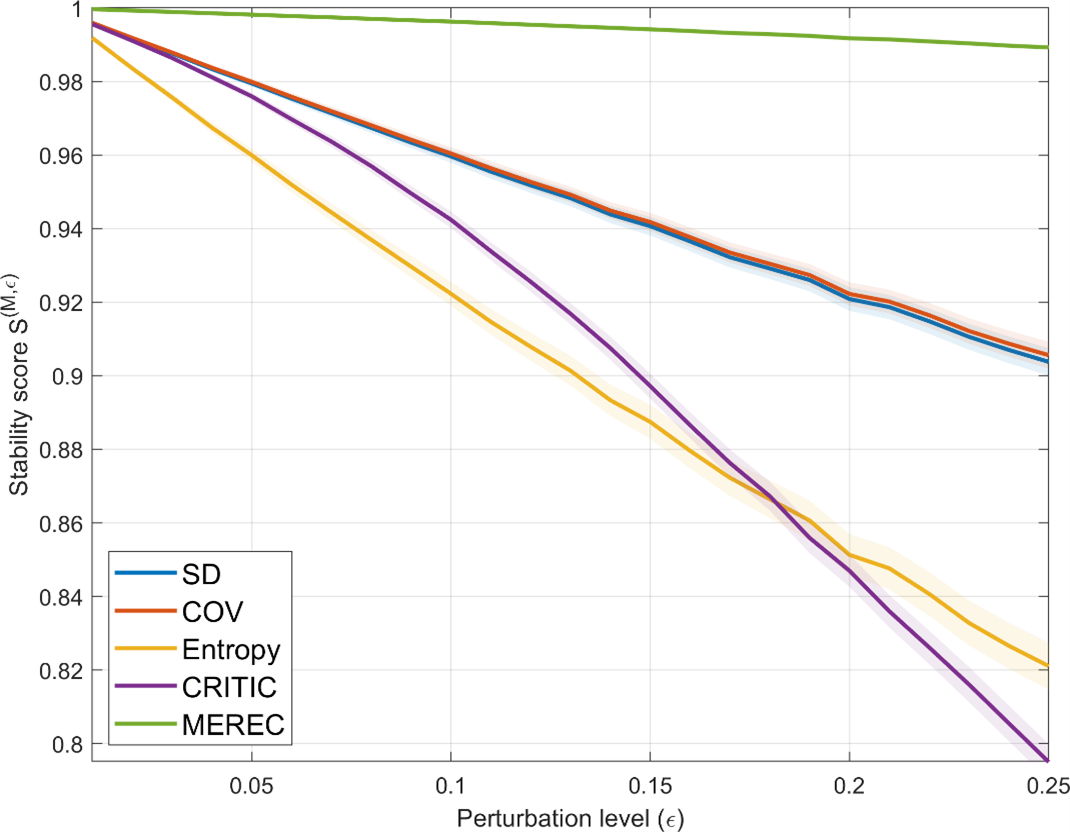}
	\caption{Stability scores $S^{(M,\varepsilon)}$ of weighting methods under vector normalization for perturbation levels $\varepsilon$. Shaded areas represent 95\% confidence intervals from Monte Carlo simulations}
	\label{fig:stability_scores}
\end{figure}

As we can observe from Fig.~\ref{fig:stability_scores}, the superiority of vector normalization is especially clear in its consistent performance across different perturbation levels, with SD, COV, and MEREC ensuring stability scores above 0.90 at $\varepsilon \leq 0.20$, and especially the MEREC exhibits remarkable resilience with scores consistently above 0.98 even at $\varepsilon = 0.25$ (0.98922). In contrast, under linear scale transformation, we noticed that only SD and COV maintain scores above 0.90 at $\varepsilon = 0.20$ (0.91882 and 0.93062, respectively), while MEREC decreases to 0.8188 and shows the worst performance among the examined methods. Under min-max normalization, although SD and COV show similar robustness, MEREC yields mathematically undefined results due to the incompatibility of its logarithmic transformation with the [0,1] bounded domain produced by min-max normalization.

Moreover, we conducted a comparative analysis between the original MEREC normalization and vector normalization to validate our methodology reasonably. While implementing MEREC following its linear scale normalized formulation, the linear approach produced extreme instability. Our findings indicated that the performance of the MEREC is considerably enhanced due to vector normalization, as shown through a mean stability score of 0.9948 against 0.9226, and the variance was reduced up to 145 times from 1.4447$\times$10$^{-3}$ to 9.9518$\times$10$^{-6}$. The improvement in stability is further observed at a high level of perturbation ($\varepsilon = 0.25$), where the vector-normalized MEREC performs with a stability score of 0.98922 compared to the score of 0.86586 for the original formulation, demonstrating a 7.82\% enhancement in mean stability in our case study.

\subsubsection{Weight distribution analysis}
\begin{figure}[htbp]
	\centering
	\includegraphics[width=\textwidth]{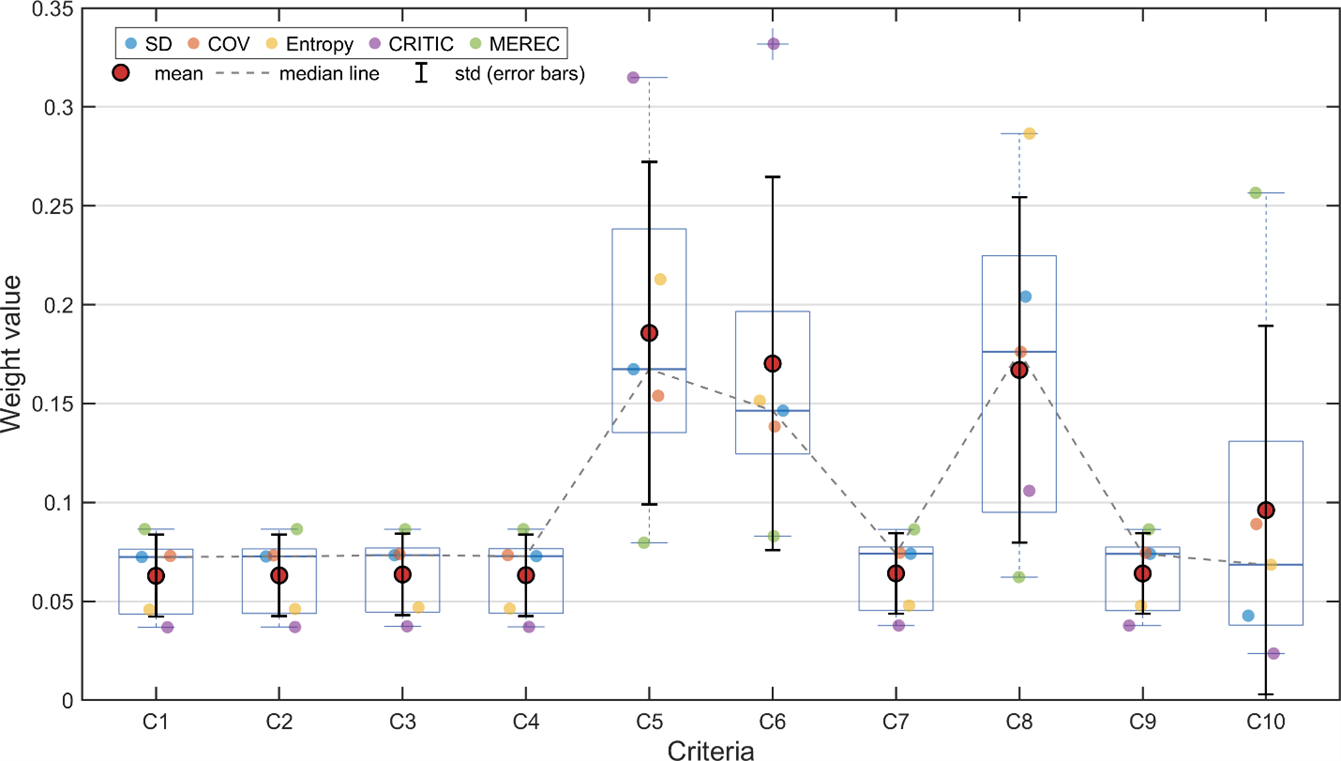}
	\caption{Distribution of criteria weights across the five weighting methods}
	\label{fig:weight_distribution}
\end{figure}

All five objective weighting methods were employed to determine the criteria weights, as presented in Table~\ref{tab:criteria_weights}, after verifying vector-based normalization as the most robust approach. At the same time, Figure~\ref{fig:weight_distribution} illustrates the weight distributions through boxplots, highlighting various statistics, including mean values indicated by red dots, median lines represented as dashed lines, and standard deviation bars, in addition to colored markers representing individual method weights for each criterion.

\begin{table}
	\caption{Criteria weights obtained by different objective weighting methods}
	\label{tab:criteria_weights}
	\centering
	\begin{tabular}{llllll}
		\toprule
		Criteria & SD & COV & Entropy & CRITIC & MEREC \\
		\midrule
		C1 & 0.0724 & 0.0730 & 0.0458 & 0.0369 & 0.0865 \\
		C2 & 0.0727 & 0.0732 & 0.0461 & 0.0370 & 0.0865 \\
		C3 & 0.0733 & 0.0739 & 0.0469 & 0.0374 & 0.0864 \\
		C4 & 0.0728 & 0.0734 & 0.0463 & 0.0371 & 0.0865 \\
		C5 & 0.1674 & 0.1539 & 0.2128 & 0.3148 & 0.0796 \\
		C6 & 0.1464 & 0.1384 & 0.1515 & 0.3319 & 0.0829 \\
		C7 & 0.0741 & 0.0746 & 0.0478 & 0.0378 & 0.0864 \\
		C8 & 0.2041 & 0.1762 & 0.2865 & 0.1059 & 0.0623 \\
		C9 & 0.0740 & 0.0745 & 0.0478 & 0.0377 & 0.0864 \\
		C10 & 0.0428 & 0.0891 & 0.0685 & 0.0236 & 0.2566 \\
		\bottomrule
		max/min & 4.7736 & 2.4138 & 6.2586 & 14.0570 & 4.1207 \\
	\end{tabular}
\end{table}

The results were compared, identifying three types of dependencies in grouping the criteria sets. All methods indicated agreement on the role of the environmental criteria (C1-C4), with average weights of approximately 0.07 and minimal standard deviations. This consistency is more noticeable, as shown by the compact boxplots and closely spaced scatter plot markers for these criteria in Fig.~\ref{fig:weight_distribution}, indicating agreement among the methods concerning the significance of environmental criteria in the sustainability evaluation process.

Cost-related criteria (C5-C8), however, were the ones that showed the greatest dispersion, as demonstrated by the notably larger boxplots and the wider standard deviation bars in Fig.~\ref{fig:weight_distribution}. The CRITIC method demonstrated significant differentiation, assigning notably high weights to Material Cost (C5 = 0.3148) and Energy Cost (C6 = 0.3319) which is identified as outlier in Fig.~\ref{fig:weight_distribution}, resulting in a max/min ratio of 14.057. The Entropy method also placed cost criteria second to End-of-Life Cost (C8 = 0.2865), resulting in a max/min ratio of 6.2586. The COV method maintained a slight advantage in the cost distributions, resulting in the lowest max/min ratio (2.4138) among all methods, indicative of an overall better balance while still allowing discrimination.

The performance criteria C9 and C10 showed the most interesting contrast. Like environmental criteria, the mass (C9) exhibited remarkable uniformity, whereas the specific stiffness (C10) demonstrated considerable variability. This divergence is mainly due to MEREC's significant emphasis on C10 (0.2566) compared to CRITIC's minimal emphasis (0.0236). The SD method exhibited a marginal overall differentiation (max/min ratio = 4.7736); within these weights, the overall median values across the entire criteria categories maintained relative consistency.

Therefore, we observe a significant variation in criteria weights across the five objective methods presented in Table~\ref{tab:criteria_weights}, as max/min ratios (ranging from 2.4138 to 14.057), emphasizing the need to use various weighting methods to ensure robust decision-making. The MEREC method achieves a relative balance of the environmental criteria while maintaining meaningful differentiation in performance criteria, especially for specific stiffness (C10). The SD and COV methods behaved similarly in weight distribution; however, the CRITIC and Entropy methods exhibited more extreme weight assignment, particularly in cost-related criteria, potentially introducing bias in the final sustainability assessment.

The weight distribution studies provide additional insights that are significant with respect to the stability analysis. More balanced weighting methods (SD, COV, MEREC) demonstrated better stability characteristics, while methods demonstrating extreme weightings (CRITIC and Entropy) were less stable. This relationship between the balance of weights and the stability of the methods strongly suggests that these issues are related, and this needs to be examined further using hierarchical clustering analysis to confirm the choice of method. The following subsection describes the findings of a hierarchical clustering study used to evaluate these observations and identify the weight methods that work best together.


\subsubsection{Hierarchical clustering analysis}
We applied hierarchical cluster analysis based on Ward's minimum variance method to confirm and classify the weighting methods according to their statistical characteristics~\cite{Murtagh2014}. Our implementation included the main features based on the stability analysis results presented in Section 4.2.1, the mean value of the stability score, and their variances for all perturbation levels. The number of clusters was established through cluster dendrogram structure and confirmed by silhouette scores, which depict the extent to which an object belongs to its cluster concerning the other clusters~\cite{Kimes2017}.

\begin{figure}[h!]
	\centering
	\includegraphics[width=0.8\textwidth]{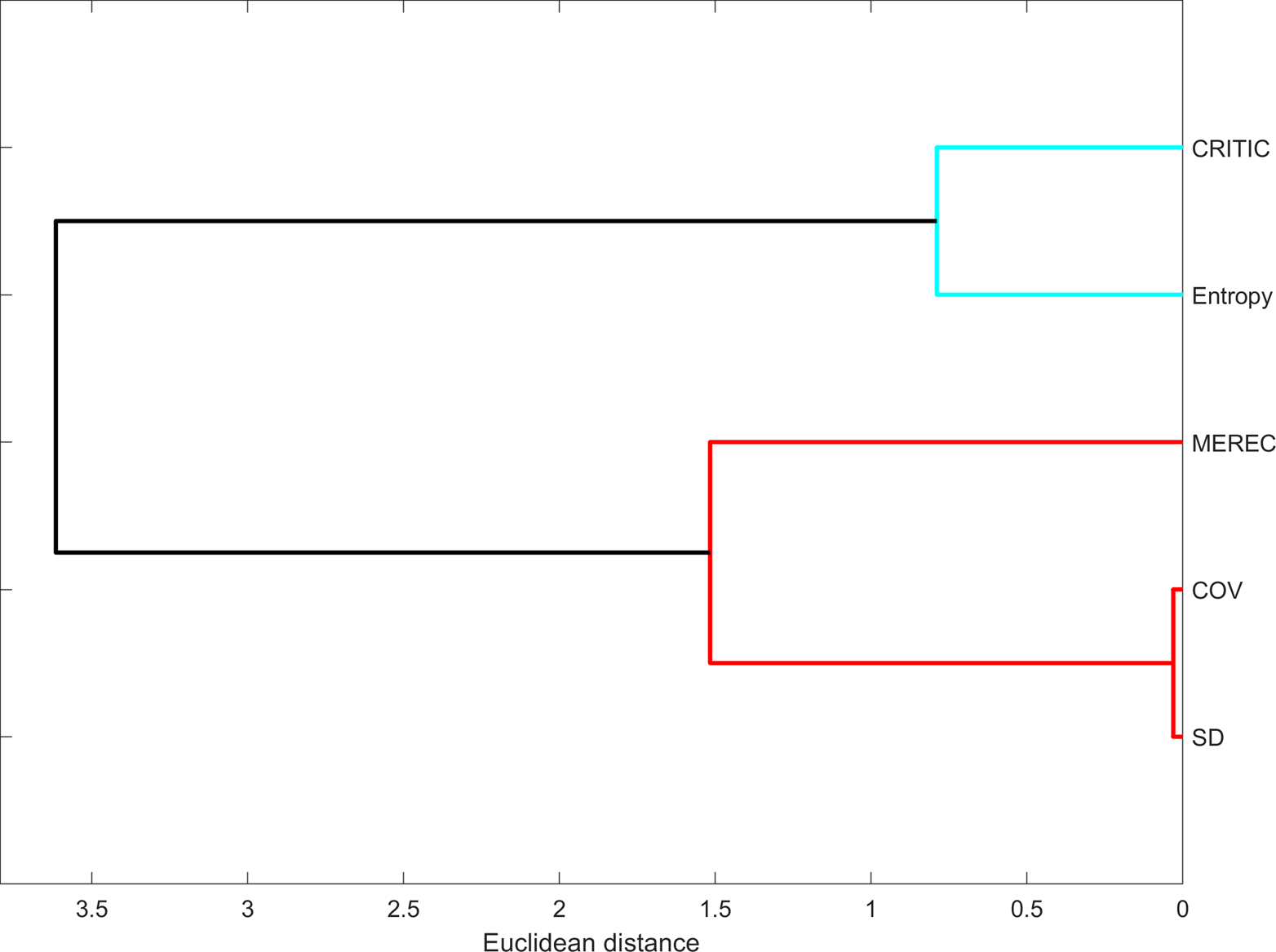}
	\caption{Hierarchical clustering dendrogram of the weighting methods based on mean stability scores and their variances across all perturbation levels}
	\label{fig:dendrogram}
\end{figure}

The dendrogram in Figure~\ref{fig:dendrogram} reveals distinct method groupings relative to stability performance. Interestingly, the SD and COV methods were closely grouped at a small Euclidean distance of less than 0.5, suggesting nearly identical stability measures since both methods approach a mean stability of 0.9491 with a very low range of 1e-6. The MEREC method, although located in a distinct cluster, has achieved a better stability measure (mean=0.9948, variance=1e-5) and joined the group of SD-COV at a considerable distance (approximately 1.5). On the contrary, Entropy and CRITIC methods exhibited considerable dispersion and were closer to each other but at a greater distance ($>$3.0), with much lower stability scores of 0.9026 and 0.9083, respectively, and higher variances, indicating increased sensitivity to perturbations in the decision matrix.

The hierarchical clustering results strongly support the further selection of MEREC, SD, and COV for weight aggregation. This cluster analysis reveals that these three methods have distinctive but complementary features; MEREC has the highest average stability (mean=0.9948) among its nearest clusters, whereas SD and COV have minimal variance (1e-6), making them extremely stable. Their distinct separation from the relatively less stable methods of Entropy and CRITIC in the dendrogram and their moderate weight proportions 4.7736, 2.4138, and 4.1207 for SD, MEREC, and COV, respectively, show that these methods will be reliable for sustainability assessment. Their high stability and the fact that they have complementary weight patterns make them appropriate for their selection for the rank analysis that follows.


\subsection{Rank analysis using aggregated weights}
After conducting validation based on hierarchical clustering, the three most stable weighting methods (SD, COV, and MEREC) were selected for the aggregation of weights using the geometric mean (GM) \cite{Aczel1983} as it preserves ratio scale properties and at the same time minimizing the influence of outlier values. The weight distributions across the three selected methods and their geometric mean values are presented in Table \ref{tab:gm_weights}.

\begin{table}[htbp]
	\caption{Weight distribution across three selected methods and their geometric mean (GM)} 
	\label{tab:gm_weights}
	\begin{center}       
		\begin{tabular}{lllll}
			\toprule
			Criterion & SD & COV & MEREC & GM \\
			\midrule
			C1 & 0.0724 & 0.0730 & 0.0865 & 0.0818\\
			C2 & 0.0727 & 0.0732 & 0.0865 & 0.0820\\  
			C3 & 0.0733 & 0.0739 & 0.0864 & 0.0825\\
			C4 & 0.0728 & 0.0734 & 0.0865 & 0.0821\\
			C5 & 0.1674 & 0.1539 & 0.0796 & 0.1350\\
			C6 & 0.1464 & 0.1384 & 0.0829 & 0.1263\\
			C7 & 0.0741 & 0.0746 & 0.0864 & 0.0830\\
			C8 & 0.2041 & 0.1762 & 0.0623 & 0.1390\\
			C9 & 0.0740 & 0.0745 & 0.0864 & 0.0830\\  
			C10 & 0.0428 & 0.0891 & 0.2566 & 0.1054\\
			\midrule
			max/min & 4.7736 & 2.4138 & 4.1207 & 1.6979\\
			\bottomrule
		\end{tabular}
	\end{center}
\end{table}

The aggregated weights show an improved balance across criteria with the max/min 1.6979, significantly lower than that noted in individual methods (SD:4.7736, COV: 2.4138, MEREC: 4.1207). This reduced dispersion shows that extreme weightings were moderate while allowing effective differentiation among criteria, specifically environmental criteria (C1-C4: $\approx$ 0.082 ), cost (C5-C8: 0.126-0.139), and performance (C9-C10: 0.083-0.105) were distinctly assessed.

\begin{table}[h!] 
	\caption{Categorical weight distribution for different weighting methods}
	\label{tab:categorical_weights}
	\begin{center}
		\begin{tabular}{lcccc}
			\toprule
			Category & SD & COV & MEREC & GM\\
			\midrule
			Environmental (C1-C4) & 29.13\% & 29.34\% & 34.59\% & 32.85\%\\
			Cost (C5-C8) & 59.19\% & 54.30\% & 31.12\% & 48.32\%\\
			Performance (C9-C10) & 11.68\% & 16.36\% & 34.29\% & 18.83\%\\  
			\midrule
			max/min & 5.07 & 3.32 & 1.11 & 2.57\\
			\bottomrule  
		\end{tabular}
	\end{center}
\end{table}

\begin{figure}[ht]
	\centering
	\includegraphics[width=0.9\textwidth]{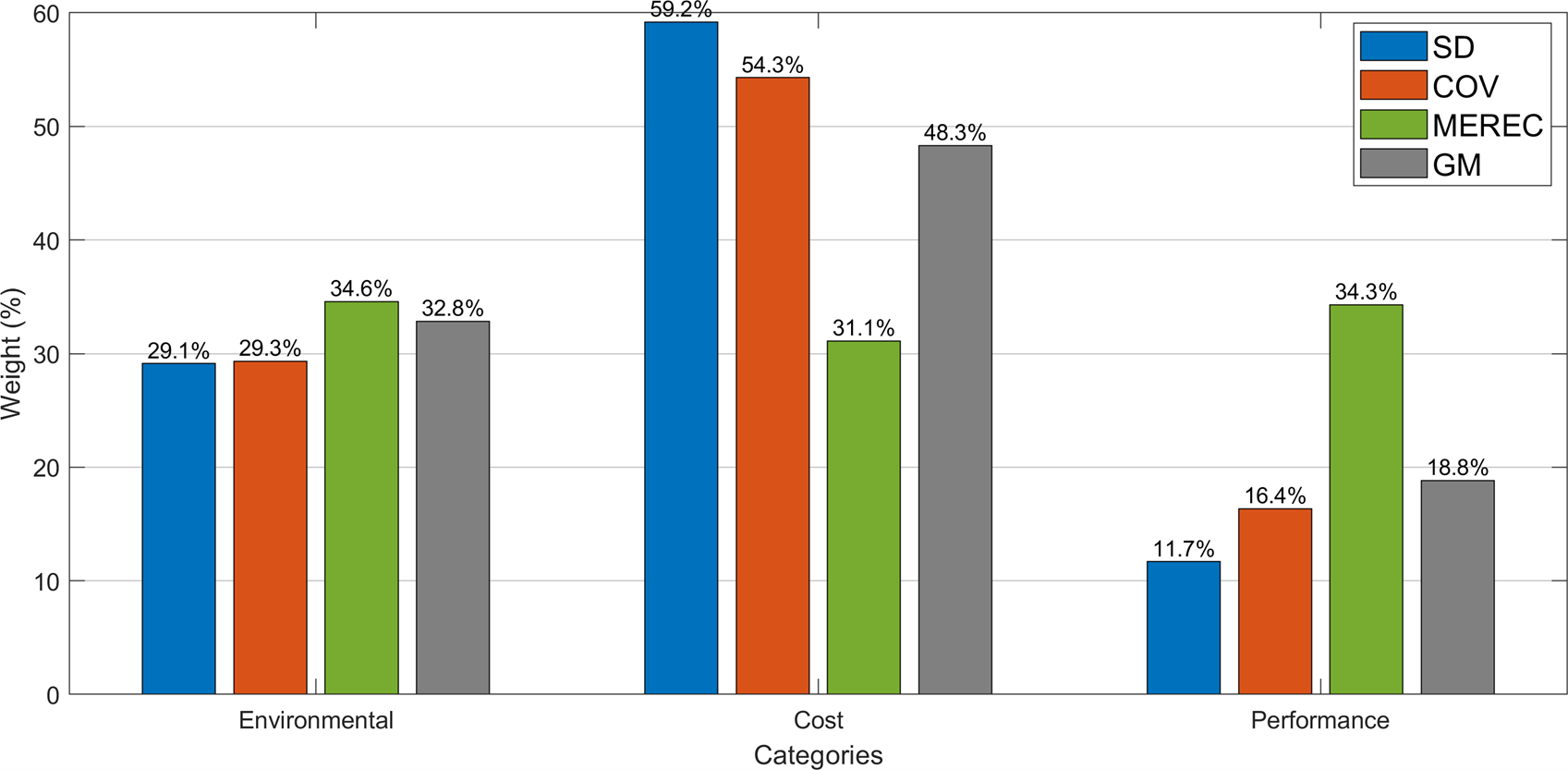} 
	\caption{Comparative weight distribution across categories}\label{fig:cat_weights}
\end{figure}

In order to elaborate on the weight distribution further, categorical analysis was performed (Table \ref{tab:categorical_weights} and Figure \ref{fig:cat_weights}). The results reveal that SD and COV are focused more on cost criteria, assigning 59.19\% and 54.30\%, respectively, but in the case of MEREC, categorization is not as extreme with environmental, cost, and performance, having percentages of 34.59\%, 31.12\%, and 34.29\% respectively. The geometric mean effectively moderates these extremes, yielding a distribution of 32.85\% for environmental, 48.32\% for cost, and 18.83\% for performance criteria. This moderation is reflected in the max/min ratio 2.57 , which falls between the highly skewed SD ratio of 5.07 and the extremely balanced MEREC ratio of 1.11.

\begin{figure}[ht]
	\centering
	\includegraphics[width=0.8\textwidth]{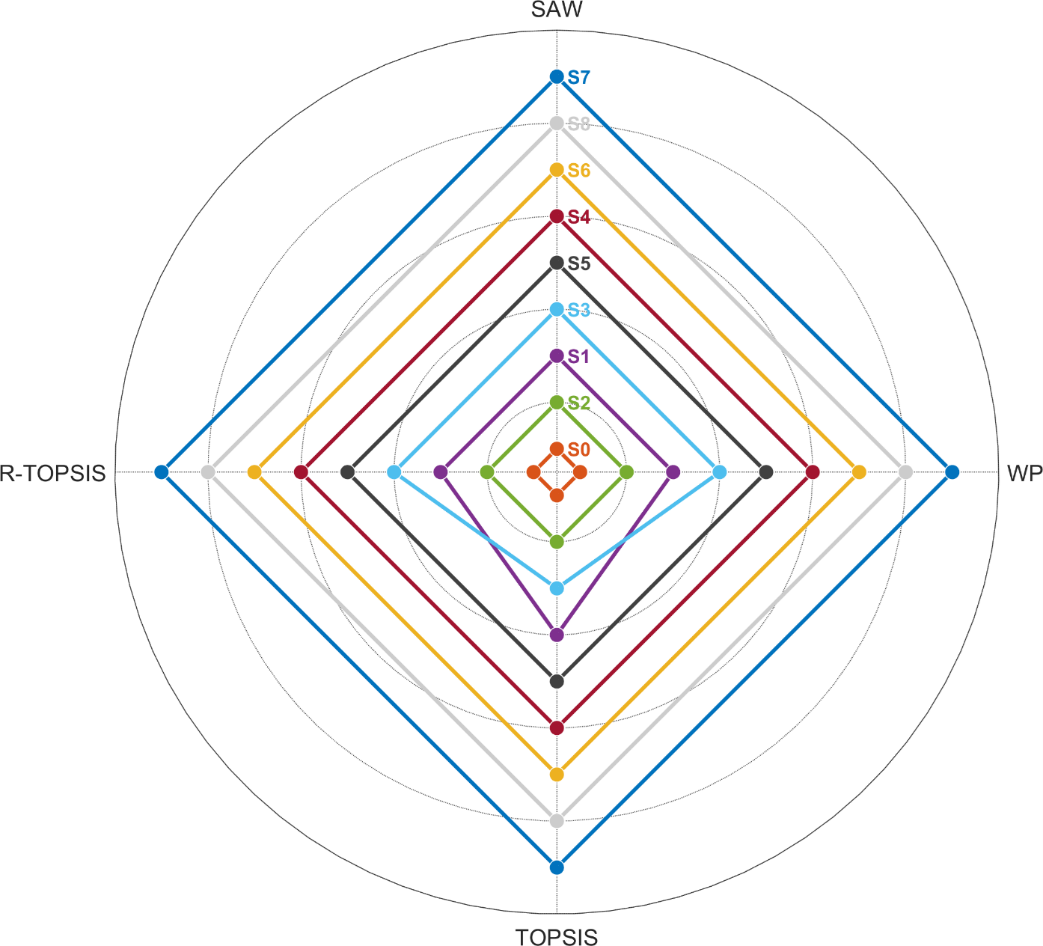} 
	\caption{Rankings comparison across the four MCDM methods (SAW, WP, TOPSIS, R-TOPSIS)}\label{fig:polar}
\end{figure}

To verify the ranking results thoroughly, these aggregated weights were implemented through four MCDM methods: SAW, WP, TOPSIS, and R-TOPSIS. The ranking results based on the obtained scores are displayed in Table \ref{tab:rank_results}, and there is a high degree of agreement among all the employed methodologies. Panel S7 (CFRP Thermoplastic) was the top alternative across all the methods and achieved the maximum scores (SAW: 0.74463, WP: 1, TOPSIS: 0.69506, R-TOPSIS: 0.73806 ). S8 and S6 (also thermoplastic CFRP) were second and third in the ranking without deviation, while the reference aluminum panel (S0) was always last. The polar plot in Fig. \ref{fig:polar} assists in presenting the ranking patterns that were reported in Table \ref{tab:rank_results} across all methods.  

\begin{table}[htbp]
	\caption{Rank analysis results across different MCDM methods using GM weights.} 
	\label{tab:rank_results}
	\begin{center}
		\begin{tabular}{lcccccc}
			\toprule
			Panel & SAW & WP & TOPSIS & R-TOPSIS & Mean & Rank\\
			\midrule
			S0 & 9 & 9 & 9 & 9 & 9.00 & 9\\
			S1 & 7 & 7 & \textbf{6} & 7 & 6.75 & 7\\ 
			S2 & 8 & 8 & 8 & 8 & 8.00 & 8\\
			S3 & 6 & 6 & \textbf{7} & 6 & 6.25 & 6\\
			S4 & 4 & 4 & 4 & 4 & 4.00 & 4\\
			S5 & 5 & 5 & 5 & 5 & 5.00 & 5\\
			S6 & 3 & 3 & 3 & 3 & 3.00 & 3\\
			S7 & 1 & 1 & 1 & 1 & 1.00 & 1\\
			S8 & 2 & 2 & 2 & 2 & 2.00 & 2\\ 
			\bottomrule
		\end{tabular}
	\end{center}
\end{table}

The remarkable consistency across different MCDM methodologies, supported by clear visual patterns in Fig. \ref{fig:polar}, validates the ranking results. The analysis conclusively identifies thermoplastic CFRP panels, particularly alternative S7, as the optimal choice when considering the complete spectrum of environmental, economic, and performance criteria.  

The comprehensive validation using multiple MCDM methods, together with the consistency of results, strongly demonstrates that the methodology and the resulting rankings are reliable. The balanced weight distribution achieved through geometric mean aggregation has contributed to stable and meaningful rankings across all methodologies.

\section{Conclusions}
In this paper, we introduced a robust sustainability assessment methodology tailored for evaluating aircraft components. This methodology integrates decision-making analysis across ten criteria, categorized into environmental impact, cost, and performance. The robustness of the approach is enhanced through the use of multiple normalization methods for the decision matrix, various objective weighting methods for the criteria, and diverse MCDM (Multi-Criteria Decision-Making) techniques for ranking alternatives. Additionally, robustness and statistical analyses were performed on the weighting methods to derive a refined and stable weighting matrix.

The methodology was applied to a fuselage panel made from either aluminum or CFRP (Carbon Fiber Reinforced Polymer). Nine design alternatives were generated by varying materials, joining methods, and component thickness. The proposed development can connect the impact on the environment, costs, and the aircraft's performance in a single measure through a multi-criteria approach.

The comprehensive analysis of the results revealed several key findings:

\begin{enumerate}
	\item The application of specific software packages e.g., SimaPro along with the Ecoinvent 3 database, supported environmental and economic analyses (LCA and LCC) as well as ANSYS to evaluate the structural behavior of the component, which allows assessing all the sustainability criteria of the component throughout its life cycle on environmental, economic, and technical perspectives. Such systematic methods, with the use of established industry standards, enhance the validity of the data and the results obtained, which is very important in enhancing the validity of sustainability assessment in aviation.
	
	\item Utilizing numerous MCDM methods and objective weighting methods, we were able to show that:
	\begin{itemize}
		\item The geometric mean of weights of the three most stable methods, that is, SD, COV, and MEREC, results in a more balanced set of weights with a max/min ratio of 1.6979, which prevents the assessment from being driven by any one aspect of sustainability. Moreover, the categorical analysis of weight distributions shows that traditional methods (SD and COV) tend to prioritize the cost category of the criteria set (54-59\%), 
		but on the other hand, the MEREC suggests a more balanced distribution across environmental (34.59\%), cost (31.12\%) and performance (34.29\%) categories. Geometric mean somehow reduces these extremes and provide a more reasonable approach to making design decisions from the sustainability perspective.
		
		\item The methodology shows notable robustness through consistent ranking results obtained by applying four different MCDM methods (SAW, WP, TOPSIS, and R-TOPSIS). This consistency validates not only the aggregation of weights but also confirms the reliability of the overall assessment framework. Remarkably, thermoplastic CFRP panel configuration S7 was the best option in all the methods applied, followed by the S8 panel configuration and S6, while the reference aluminum panel S0 always ranked last.
	\end{itemize}
\end{enumerate}

In addition, the methodology developed in this study is fully parametric, making it applicable to any structural component. It can also seamlessly integrate with optimization modules to support sustainability-driven, eco-design approaches for aircraft components.

\section*{Acknowledgement}
\addcontentsline{toc}{section}{Acknowledgement}
The work described in the paper has been financially supported by the Clean Aviation project FASTER-H2 (Project: 101101978). The views and opinions expressed in the paper are those of the author(s) only and do not necessarily reflect those of the European Union or Clean Aviation Joint Undertaking.

\addcontentsline{toc}{section}{References}

\appendix

\section*{Appendix A1. Materials}
\addcontentsline{toc}{section}{Appendix A1. Materials}

\begin{center}
	\textit{Table A1: Composition of Al2024 (Density: 2780 kg/m3)} \\
	\begin{tabular}{|l|l|}
		\hline
		\textbf{Material} & \textbf{Content} \\
		\hline
		Aluminum & $92.81 \%$ \\
		\hline
		Chromium & $0.05 \%$ \\
		\hline
		Iron & $0.25 \%$ \\
		\hline
		Magnesium & $0.5 \%$ \\
		\hline
		Manganese & $0.6 \%$ \\
		\hline
		Silicon & $0.25 \%$ \\
		\hline
		Titanium & $0.07 \%$ \\
		\hline
		Zinc & $0.125 \%$ \\
		\hline
		Copper & $4.36 \%$ \\
		\hline
	\end{tabular}
\end{center}

\begin{center}
	\textit{Table A2: Composition of Al7075 (Density: 2810 kg/m3)} \\
	\begin{tabular}{|l|l|}
		\hline
		\textbf{Material} & \textbf{Content} \\
		\hline
		Aluminum & $88 \%$ \\
		\hline
		Chromium & $0.2 \%$ \\
		\hline
		Iron & $0.3 \%$ \\
		\hline
		Magnesium & $2.5 \%$ \\
		\hline
		Manganese & $0.1 \%$ \\
		\hline
		Silicon & $0.35 \%$ \\
		\hline
		Titanium & $0.15 \%$ \\
		\hline
		Zinc & $6 \%$ \\
		\hline
		Copper & $1.5 \%$ \\
		\hline
	\end{tabular}
\end{center}

\begin{center}
	\textit{Table A3: Composition of CFRP thermoset material (Density: 1570 kg/m3, Fiber content: $57.7 \%$ wt.)}
	\begin{tabular}{|p{8cm}|l|}
		\hline
		\textbf{Material} & \textbf{Content} \\ 
		\hline
		Epoxy resin and Boron trifluoride hardener for 1 kg CFRP & 0.423 kg \\
		Epoxy resin for 1 kg resin & 0.66 kg \\
		Boron trifluoride for 1 kg resin & 0.33 kg \\
		\hline
		PAN carbon fibers for 1 kg thermoset CFRP & 0.577 kg \\  
		Acrylonitrile (production of 1 kg PAN carbon fibers) & 2.25 kg \\
		Dimethylacetamide (production of 1 kg PAN carbon fibers) & 0.031 kg \\
		Polyurethane (production of 1 kg PAN carbon fibers) & 0.02 kg \\
		\hline
		Embodied energy (prepreg) & 40 MJ \\ 
		Electricity (production of 1 kg PAN carbon fibers) & 58 kWh \\
		Heat (production of 1 kg PAN carbon fibers) & 257.3 MJ\\
		\hline
	\end{tabular}
\end{center}

\begin{center}
	\textit{Table A4: Composition of TP (Fiber content: 57\% wt.)}
	\begin{tabular}{|p{8cm}|l|}
		\hline
		\textbf{Material} & \textbf{Content} \\
		\hline
		Thermoplastic resin polyphenylene powder & 0.43 kg \\
		Polyphenylene sulfide (thermoplastic resin) & 1 kg \\
		\hline
		Plastic micronizer machine (milling the resin pellets) & 0.2 kWh \\
		\hline
		PAN carbon fibers for 1 kg thermoplastic CFRP & 0.57 kg \\
		\hline 
		Composite density & 1550kg/m$^3$ \\
		\hline
	\end{tabular}
\end{center}

\begin{center}
	\textit{Table A5: Bonding} \\
	\begin{tabular}{|p{8cm}|l|}
		\hline
		\textbf{Material} & \textbf{Content} \\
		\hline
		Bisphenol A epoxy-based vinyl ester resin & 0.66 kg \\
		\hline
		Ethylenediamine & 0.33 kg \\
		\hline
	\end{tabular}
\end{center}

\section*{Appendix A2. Processes}
\addcontentsline{toc}{section}{Appendix A2. Processes}

\begin{center}
	\textit{Table A6: Metal forming process} \\
	\begin{tabular}{|p{6cm}|l|}
		\hline
		\textbf{Subprocess} & \textbf{Energy (kWh/kg)} \\
		\hline
		Stretch forming & 3.95 \\
		\hline
		Incremental sheet forming & 12.91 \\
		\hline
		Hydroforming & 4 \\
		\hline
		Hot rolling mills & 0.07 \\
		\hline
	\end{tabular}
\end{center}

\begin{center}
	\textit{Table A7: Friction stir welding energy} \\
	\begin{tabular}{|p{6cm}|l|}
		\hline
		\textbf{Process} & \textbf{Energy (kwh/800mm)} \\
		\hline
		Friction stir welding & 0.07056 \\
		\hline
	\end{tabular}
\end{center}

\begin{center}
	\textit{Table A8: Autoclave process energy} \\
	\begin{tabular}{|p{6cm}|l|}
		\hline
		\textbf{Process} & \textbf{Energy} \\
		\hline
		Vacuum generation & 10.2 MJ \\
		\hline
		Autoclave curing & 294.3 kWh \\
		\hline
	\end{tabular}
\end{center}

\section*{Appendix A3. Cost}
\addcontentsline{toc}{section}{Appendix A3. Cost}

\begin{center}
	\textit{Table A9: Autoclave process energy} \\
	\begin{tabular}{|l|l|}
		\hline
		\textbf{Material} & \textbf{Cost (€/kg)} \\
		\hline
		Al2024 & 5 \\ \hline
		Al7075 & 7 \\ \hline
		Boron Trifluoride & 341.66 \\ \hline
		Epoxy resin & 9.15 \\ \hline
		PAN carbon fibers & 30 \\ \hline
		Permabond adhesive & 693.5 \\ \hline
		Polyphenylene sulfide & 20 \\ \hline
		Polyurethane sizing & 3.58 \\ 
		\hline
	\end{tabular}
\end{center}

\begin{center}
	\textit{Table A10: Electricity} \\
	\begin{tabular}{|l|l|}
		\hline
		\textbf{Energy} & \textbf{Cost (€/kWh)} \\
		\hline
		Electricity Europe & 0.2847 \\
		\hline
	\end{tabular}
\end{center}

\section*{Appendix A4. Use-phase of the panel}
\addcontentsline{toc}{section}{Appendix A4. Use-phase of the panel}

\begin{center}
	\textit{Table A11: Fuel consumption (Airbus A319)} \\
	\begin{tabular}{|l|l|}
		\hline
		Total distance in A319 lifetime & 20188120.8 km \\ \hline
		Kerosene consumption per kg of transportation & 6729.374 kg kerosene/ kg transportation \\ \hline
		Kerosene cost & 0.658 €/kg \\
		\hline
	\end{tabular}
\end{center}

\section*{Appendix A5. End-of-life of the panel}
\addcontentsline{toc}{section}{Appendix A5. End-of-life of the panel}

\begin{center}
	\textit{Table A12: Waste scenarios} \\
	\begin{tabular}{|l|l|}
		\hline
		Aluminum panel & 85\% recycling and 15\% landfill \\ \hline
		CFRP panel & 100\% landfill \\
		\hline
	\end{tabular}
\end{center}

\begin{center}
	\textit{Table A13: End-of-life cost} \\
	\begin{tabular}{|l|l|}
		\hline
		Landfill cost & 0.06 €/kg \\ \hline
		Recycling cost & 1.04 €/kg \\
		\hline
	\end{tabular}
\end{center}

\end{document}